\documentclass[english]{scrartcl}
\usepackage{lmodern}

\usepackage[T1]{fontenc}
\usepackage[latin9]{inputenc}
\usepackage{geometry}
\usepackage{color}
\usepackage{babel}
\usepackage{array}
\usepackage{float}
\usepackage{url}
\usepackage{amsmath}
\usepackage{amsthm}
\usepackage{amssymb}
\usepackage{stmaryrd}
\usepackage{graphicx}
\usepackage[authoryear]{natbib}
\usepackage[unicode=true,pdfusetitle,
 bookmarks=true,bookmarksnumbered=false,bookmarksopen=false,
 breaklinks=false,pdfborder={0 0 0},pdfborderstyle={},backref=false,colorlinks=true]
 {hyperref}
\hypersetup{
 citecolor=blue}

\makeatletter

\providecommand{\tabularnewline}{\\}
\floatstyle{ruled}
\newfloat{algorithm}{tbp}{loa}
\providecommand{\algorithmname}{Algorithm}
\floatname{algorithm}{\protect\algorithmname}

\newenvironment{lyxcode}
	{\par\begin{list}{}{
		\setlength{\rightmargin}{\leftmargin}
		\setlength{\listparindent}{0pt}
		\raggedright
		\setlength{\itemsep}{0pt}
		\setlength{\parsep}{0pt}
		\normalfont\ttfamily}%
	 \item[]}
	{\end{list}}
\theoremstyle{plain}
\newtheorem{thm}{\protect\theoremname}
\theoremstyle{plain}
\newtheorem*{cor*}{\protect\corollaryname}

\usepackage[font={footnotesize,it},labelfont=bf,hangindent=-2cm]{caption}

\makeatother

\providecommand{\corollaryname}{Corollary}
\providecommand{\theoremname}{Theorem}

\begin{document}
\title{Dynamic programming by polymorphic semiring algebraic shortcut fusion}
\author{Max A. Little$^{\star,\dagger}$, Xi He$^{\star}$, Ugur Kayas$^{\#}$}
\maketitle
\begin{center}
\emph{$^{\star}$School of Computer Science, University of Birmingham,
UK}
\par\end{center}

\begin{center}
\emph{$^{\dagger}$MIT, Cambridge, MA, USA}
\par\end{center}

\begin{center}
\emph{$^{\#}$Abdullah Gul University, Turkey}
\par\end{center}

\begin{center}
\emph{First author contact: maxl@mit.edu}\footnote{\emph{This work partially funded by NIH grant UR-Udall Center, award
number P50 NS108676.}}
\par\end{center}
\begin{abstract}
Dynamic programming (DP) is a broadly applicable algorithmic design
paradigm for the efficient, exact solution of otherwise intractable,
combinatorial problems. However, the design of such algorithms is
often presented informally in an ad-hoc manner. It is sometimes difficult
to justify the correctness of these DP algorithms. To address this
issue, this paper presents a rigorous algebraic formalism for systematically
deriving DP algorithms, based on semiring polymorphism. We start with
a specification, construct a (brute-force) algorithm to compute the
required solution which is self-evidently correct because it exhaustively
generates and evaluates all possible solutions meeting the specification.
We then derive, primarily through the use of shortcut fusion, an implementation
of this algorithm which is both efficient and correct. We also demonstrate
how, with the use of semiring lifting, the specification can be augmented
with combinatorial constraints and through semiring lifting, show
how these constraints can also be fused with the derived algorithm.
This paper furthermore demonstrates how existing DP algorithms for
a given combinatorial problem can be abstracted from their original
context and re-purposed to solve other combinatorial problems.

This approach can be applied to the full scope of combinatorial problems
expressible in terms of semirings. This includes, for example: optimization,
optimal probability and Viterbi decoding, probabilistic marginalization,
logical inference, fuzzy sets, differentiable softmax, and relational
and provenance queries. The approach, building on many ideas from
the existing literature on constructive algorithmics, exploits generic
properties of (semiring) polymorphic functions, tupling and formal
sums (lifting), and algebraic simplifications arising from constraint
algebras. We demonstrate the effectiveness of this formalism for some
example applications arising in signal processing, bioinformatics
and reliability engineering. Python software implementing these algorithms
can be downloaded from: \url{http://www.maxlittle.net/software/dppolyalg.zip}.
\end{abstract}

\section{Introduction}

\emph{Dynamic programming }(DP) \citep{bellman-1957,Sniedovich-2011}
is one of the most effective and widely used computational tools for
finding exact solutions to a large range of otherwise intractable
combinatorial problems \citep{kleinberg-2005}. Typically, the exhaustive
(brute-force) solution to problems for which DP is amenable are of
exponential or even factorial, time complexity. Where DP is applicable,
it is often possible to reduce the worst case computational effort
required to solve the problem, to something tractable such as low-order
(quasi)-polynomial.

Nonetheless, devising correct and efficient DP algorithms typically
relies on special intuition and insight \citep{demoor-1999}. It is
also often difficult to prove that an algorithm is correct with respect
to a formal problem specification and gain understanding of the function
of these algorithms from their, sometimes inscrutable, implementations.
To address these shortcomings, a more systematic approach is to start
with a specification of the combinatorial problem and then \emph{compute}
an efficient implementation of the same, through provably correct
\emph{derivation }steps. In this way, the resulting algorithm is both
efficient and guaranteed correct. This approach is exemplified in
\emph{constructive algorithmics} frameworks described in e.g. \citet{bird-1996},
\citet{deMoor-1991} and \citet{Jeuring-1993}.

This paper reports on a novel and relatively simple approach to the
systematic design of such DP algorithms which exploits semiring polymorphism
and shortcut fusion. Our approach starts from a specification given
in terms of a \emph{semiring objective }over the \emph{set of all
combinatorial configurations }in the DP problem. With a semiring polymorphic
generator of this set of combinatorial configurations (often given
by a Bellman recursion) we show how to derive an efficient and correct
algorithm for solving the semiring objective by applying semiring
\emph{shortcut fusion}. We furthermore show how this approach can
be extended to complex DP problems which solve semiring objectives
over multiple \emph{combinatorial constraints}.

Our approach brings together several concepts that have traditionally
existed independently across various fields, including machine learning,
computational linguistics, and automata theory. \emph{Semirings} \citep{golan-1999}
are widely used in special DP applications \citep{huang-2008,goodman-1999,mensch-2018,li-2009},
but their general usage in DP lacks rigorous justifications in terms
of correctness with respect to a specification, which we provide here.
We furthermore show how \emph{semiring lifting} \citep{emoto-2012}
can be used to solve more complex \emph{constrained }DP problems where
the constraint is expressed in terms of an algebra homomorphism. When
this constraint algebra is a \emph{group} we show that it further
simplifies the algorithm implementation. We also show how semirings
can often be \emph{combined in parallel }(\emph{tupled}) to significant
advantage such as providing a general implementation of backtracing
applicable to any such semiring DP algorithm.

Our approach is applicable to a very wide range of unconstrained or
constrained combinatorial problems which can be expressed in terms
of semirings. We demonstrate its effectiveness on some practical,
novel extensions of classical problems in signal processing, machine
learning, computational statistics and engineering.

The layout of the paper is as follows. In Section \ref{sec:Theory},
we detail the main theoretical developments of this paper, and in
Section \ref{sec:Applications} we develop DP algorithms for applications
from several disciplines. Section \ref{sec:Related-work} puts the
work into the context of existing research on DP algorithms in general.
We end in Section \ref{sec:Discussion-and-conclusions} with a summary
and discussion of the importance, general scope and possible extensions
of the work. The appendices contain detailed proofs of the main results
in the paper, list some widely-used semirings and simplified constraint
algebras.

\section{Theory\label{sec:Theory}}

In this paper, sets are indicated by the upper case double-strike
letters $\mathbb{S},\mathbb{T}$ with their corresponding cardinalities,
$S=\left|\mathbb{S}\right|$ and $T=\left|\mathbb{T}\right|$, or
the standard sets $\mathbb{R},\mathbb{N}$ etc. The Boolean set is
given by $\mathbb{B}=\left\{ T,F\right\} $ (for true, false respectively).
To indicate the type of \emph{sets with elements of type} $\mathbb{S}$,
we write $\left\{ \mathbb{S}\right\} $. The type of \emph{lists }(that
is, ordered sequences) with \emph{elements in the arbitrary type }$\mathbb{S}$
are denoted with $\left[\mathbb{S}\right]$. Binary algebraic operators
are written as circled symbols, $\oplus,\otimes,\odot$ and their
corresponding identities are $i_{\oplus},i_{\otimes,}i_{\odot}$.
Algebras and objects such as monoids, groups and semirings using binary
operators are given as tuples with upper-case calligraphic letters
for names, e.g. $\mathcal{S}=\left(\mathbb{S},\oplus,\otimes,i_{\oplus},i_{\otimes}\right)$
is a semiring over values of type $\mathbb{S}$ with binary operators
$\oplus,\otimes:\mathbb{\left(S\to\mathbb{S}\right)\to\mathbb{S}}$
with identities $i_{\oplus},i_{\otimes}\in\mathbb{S}$ and $\mathcal{M}=\left(\mathbb{M},\odot,i_{\odot}\right)$
a monoid with values of type $\mathbb{M}$ and binary operator $\odot$.
Integer and natural number indices are given by lower case letters
$n,i$ etc. We use the superscript notation $f^{u}:\mathbb{Y}$ as
shorthand for (higher-order) function application $f\left(u\right)$
which does not have any input (it is not a function). Thus, for $u$
having type $\mathbb{U}\to\mathbb{V}$ it follows the function has
type $f:\left(\mathbb{U}\to\mathbb{V}\right)\to\mathbb{Y}$. If $f$
has an additional \emph{recurrence index}, $n\in\mathbb{N}$, then
it has type $f:\left(\mathbb{U}\to\mathbb{V}\right)\to\mathbb{N}\to\mathbb{Y}$
so that $f^{u}:\mathbb{N}\to\mathbb{Y}$ is shorthand for partial
application $f\left(u\right)$, full application uses subscript notation
$f_{n}^{u}:\mathbb{Y}$. Where an algebra appears in the superscript
$f^{\mathcal{M}}$, this is shorthand for $f\left(\odot,i_{\odot}\right)$
which in this case means full application to the operators and identities
of the algebra $\mathcal{M}$. Where there is no ambiguity we suppress
superscript parameters purely for presentational clarity. Binary operators
are subscripted to indicate \emph{lifting}, e.g. $\oplus_{\mathcal{M}}$
is the $\oplus$ operator lifted over the monoid algebra $\mathcal{M}$
and the lifted semiring value $x_{m}$ denotes a function of type
$x:\mathbb{M}\text{\ensuremath{\to\mathbb{S}}}$ which has an input
of type $\mathbb{M}$ of values in the monoid algebra $\mathcal{M}$.
Regarding types, lifting means that, whereas the plain semiring operator
is a function taking two semiring values to a single semiring value
e.g. of type $\oplus:\left(\mathbb{S\to}\mathbb{S}\right)\to\mathbb{S}$,
its lifted version takes two lifted values $\mathbb{M}\text{\ensuremath{\to\mathbb{S}}}$,
returning another lifted value so it has type $\oplus_{\mathcal{M}}:\left(\left(\mathbb{M}\text{\ensuremath{\to\mathbb{S}}}\right)\to\left(\mathbb{M}\text{\ensuremath{\to\mathbb{S}}}\right)\right)\to\left(\mathbb{M}\text{\ensuremath{\to\mathbb{S}}}\right)$.

\subsection{Correct and efficient dynamic programming by semiring fusion}

A very wide class of DP problems can be specified as a \emph{combinatorial
problem }over a \emph{semiring} $\mathcal{S}=\left(\mathbb{S},\bigoplus,\bigotimes,i_{\oplus},i_{\otimes}\right)$,

\begin{equation}
\text{\ensuremath{s^{*}}=\ensuremath{\bigoplus_{l\in\mathbb{L}}}\ensuremath{\ensuremath{\bigotimes_{x\in l}}w\ensuremath{\left(x\right)}}},\label{eq:semiring-spec}
\end{equation}
where $\mathbb{L}$ is the set of all \emph{combinatorial configurations}
and $w:\mathbb{X}\to\mathbb{S}$ embeds values in the DP problem into
values in the semiring $\mathcal{S}$. As an example, consider the
\emph{rod-cutting problem} which maximizes the sum of values of \emph{segments}
of a \emph{list partition}. A partition of a list is a decomposition
of a list of values, into a list of non-empty contiguous segments.
For instance, for the list $\left[1,2,3\right]$, the set of configurations
is

\begin{equation}
\mathbb{L}=\left\{ \left[\left[1\right],\left[2\right],\left[3\right]\right],\left[\left[1,2\right],\left[3\right]\right],\left[\left[1\right],\left[2,3\right]\right],\left[\left[1,2,3\right]\right]\right\} .
\end{equation}

In this example, the combinatorial configurations are list partitions
of segments, so variable $l$ represents a partition such as $\left[\left[1\right],\left[2,3\right]\right]$
and variable $x$ a segment of that partition such as $\left[2,3\right]$,
with $w\left(x\right)$ being the value of that segment. We can specify
the rod-cutting problem as
\begin{equation}
s_{\text{rodcut}}^{*}=\max_{l\in\mathbb{L}}\underset{x\in l}{\sum}w\left(x\right).\label{eq:rod-cutting}
\end{equation}
which is an instance of (\ref{eq:semiring-spec}) with the \emph{max-plus
}semiring $\left(\mathbb{R},\max,+,-\infty,0\right)$.

We will next demonstrate how to solve (\ref{eq:semiring-spec}) through
\emph{brute-force }(\emph{exhaustive})\emph{ generate}-\emph{evaluat}e
computation. There is a special semiring which can be used to exhaustively
generate the set of all possible combinatorial configurations $\mathbb{L}$.
We call this special semiring the \emph{generator semiring} $\mathcal{G}=\left(\left\{ \left[\mathbb{X}\right]\right\} ,\cup,\circ,\emptyset,\left\{ \left[\,\right]\right\} \right)$.
This well-known semiring (and variants) arise in several contexts;
for example, to computational linguists it is called the \emph{formal
language} semiring over sets of lists with elements of type $\mathbb{X}$,
which we denote by $\left\{ \left[\mathbb{X}\right]\right\} $. The
operator $\cup$ is set union, and $l_{1}\circ l_{2}$ is the \emph{cross-join}
of two sets of lists $l_{1},l_{2}:\left\{ \left[\mathbb{X}\right]\right\} $
obtained by concatenating each element in configuration $l_{1}$ with
each element in $l_{2}$. To illustrate, for elements set $\mathbb{X}=\left\{ a,b,c,d,e\right\} $,
\begin{equation}
\left\{ \left[a,b\right],\left[c\right]\right\} \circ\left\{ \left[d\right],\left[e\right]\right\} =\left\{ \left[a,b,d\right],\left[a,b,e\right],\left[c,d\right],\left[c,e\right]\right\} .
\end{equation}

Define a \emph{semiring generator }$f^{\mathcal{S},w}:\mathbb{S}$
which is a function $f$ fully applied to the arbitrary semiring $\mathcal{S}$'s
binary operators and operator identities and \emph{value mapping }function
$w:\mathbb{X}\to\mathbb{S}$, which maps some data into semiring values.
Apply this function $f$ to the generator semiring $\mathcal{G}$
and value mapping function $w\left(x\right)=w^{\prime}\left(x\right)=\left\{ \left[x\right]\right\} $
which embeds an element $x\in\mathbb{X}$ into a \emph{singleton }configuration,
$\left\{ \left[x\right]\right\} $. In this way, we will generate
(enumerate) all combinatorial configurations by computing $\mathbb{L}=f^{\mathcal{G},w^{\prime}}$.
Also, assume we have an \emph{evaluation function} $g^{\mathcal{S},w}:\left\{ \left[\mathbb{X}\right]\right\} \to\mathbb{S}$
which is a \emph{semiring homomorphism} preserving the semiring structure
for all $l_{1},l_{2}:\left\{ \left[\mathbb{X}\right]\right\} $,

\begin{equation}
\begin{aligned}g\left(l_{1}\cup l_{2}\right) & =g\left(l_{1}\right)\oplus g\left(l_{2}\right)\\
g\left(l_{1}\circ l_{2}\right) & =g\left(l_{1}\right)\otimes g\left(l_{2}\right)\\
g\left(\emptyset\right) & =i_{\oplus}\\
g\left(\left\{ \left[\,\right]\right\} \right) & =i_{\otimes}
\end{aligned}
\label{eq:semiring-hom}
\end{equation}
along with the requirement that $g\left(\left\{ \left[x\right]\right\} \right)=w\left(x\right)$
for all $x\in\mathbb{X}$. Here, we suppress the subscripted parameters
from $g$ for clarity. Now, using this generator $f$ and evaluator
$g$, problems defined by the semiring specification (\ref{eq:semiring-spec})
can be solved exactly using the following exhaustive \emph{generate-evaluate
algorithm},

\begin{equation}
s^{*}=g^{\mathcal{S},w}\left(f^{\mathcal{G},w^{\prime}}\right).\label{eq:gen-eval}
\end{equation}

Self-evidently, this solves (\ref{eq:semiring-spec}) because $f^{\mathcal{G},w^{\prime}}$
generates the set $\mathbb{L}$ and then $g^{\mathcal{S},w}$ evaluates
each one of the configurations $l\in\mathbb{L}$ over $\otimes$ and
aggregates them over the semiring $\oplus$.

While problem (\ref{eq:semiring-spec}) is indeed correctly solved
by this generate-evaluate algorithm, this computation is usually \emph{intractable
}because the configurations set $\mathbb{L}$ is typically exponential
(or larger). However, it turns out there is no need to generate the
set $\mathbb{L}$ in the first place. We have assumed that the generator
$f$ is \emph{polymorphic in an arbitrary semiring} $\mathcal{S}$
i.e. it is a function of type
\begin{equation}
f:\left(\mathbb{S}\to\mathbb{S}\to\mathbb{S}\right)\to\left(\mathbb{S}\to\mathbb{S}\to\mathbb{S}\right)\to\mathbb{S}\to\mathbb{S}\to\left(\mathbb{X}\to\mathbb{S}\right)\to\mathbb{S}.\label{eq: polymorphic function}
\end{equation}

In words, this function takes as parameters two semiring operators
of type $\mathbb{S}\to\mathbb{S}\to\mathbb{S}$, two semiring operator
identities of type $\mathbb{S}$ and a mapping function $w:\mathbb{X\to\mathbb{S}}$,
returning a value of type $\mathbb{S}$. Then, we can show the following
equivalence holds,
\begin{equation}
s^{*}=g^{\mathcal{S},w}\left(f^{\mathcal{G},w^{\prime}}\right)=f^{\mathcal{S},w}.\label{eq:DP-semiring-fusion}
\end{equation}
provided only in addition that $g^{\mathcal{S},w}$ is a semiring
homomorphism mapping $\mathcal{G}\to\mathcal{S}$. We call this\emph{
}theorem \emph{semiring fusion}. If the semiring $\mathcal{S}$ has
operators with $O\left(1\right)$ computational complexity, then computing
$s^{*}=f^{\mathcal{S},w}$ will be much more computationally efficient
than using (\ref{eq:gen-eval}). The proof of this theorem, given
rigorously in \nameref{sec:AppendixA} is a straightforward application
of \emph{Wadler's free theorem} \citep{wadler-1989}. This theorem
is very widely applicable in the context of DP because semiring polymorphic
generators $f^{\mathcal{S},w}$ are extremely common, for instance,
all \emph{Bellman recursions }are of this form \citep{bellman-1957,Sniedovich-2011}.

\subsection{Constraint lifting}

Combinatorial problems in applications of DP, are often more complex
than those which can be specified by the simple semiring formulation
(\ref{eq:semiring-spec}). For instance, let us suppose we want to
\emph{constrain} the semiring problem to restrict it to apply only
to configurations which satisfy some constraint, $c:\left[\mathbb{X}\right]\to\mathbb{B}$.
Such constrained combinatorial problems are specified as,
\begin{equation}
s_{\mathrm{cons}}^{*}=\bigoplus_{l\in\mathbb{L}:c\left(l\right)}\underset{x\in l}{\bigotimes}w\left(x\right).\label{eq:constrained-semiring-spec}
\end{equation}

As with exhaustive generate-evaluate above, a self-evidently correct
(but not at all ``smart'') way to solve (\ref{eq:constrained-semiring-spec})
is to apply the following algorithm. Firstly, compute all configurations
$\mathbb{L}$ using a semiring generator $f^{\mathcal{G},w^{\prime}}$.
Then, remove (\emph{filter away}) those configurations $l\in\mathbb{L}$
for which $c\left(l\right)=F$ using a \emph{filtering function }$\phi^{c}:\left\{ \left[\mathbb{X}\right]\right\} \to\left\{ \left[\mathbb{X}\right]\right\} $
partially applied to the constraint function, $c$. Finally, evaluate
the remaining configurations using a evaluation function $g^{\mathcal{S},w}$.
This \emph{generate-filter-evaluate} algorithm computes,
\begin{equation}
s_{\mathrm{cons}}^{*}=g^{\mathcal{S},w}\left(\phi^{c}\left(f^{\mathcal{G},w^{\prime}}\right)\right).\label{eq:gen-filt-eval-exhaustive}
\end{equation}

The difficulty with this approach is the same as faced above: the
intractable size of the intermediate configurations $\mathbb{L}$.
Above, this was solved by invoking semiring fusion (\ref{eq:DP_fusion}),
but unfortunately here the filtering $\phi^{c}$ prevents this theorem
from being applicable. We will present a practical solution to this
which makes use of \emph{semiring lifting} \citep{Jeuring-1993,emoto-2012}.
This is based on the following idea: if we can find a new semiring
which implicitly evaluates the constraints $c$, then we can fuse
the constraint with the semiring homomorphism (\ref{eq:semiring-hom})
so that semiring fusion (\ref{eq:DP_fusion}) becomes applicable once
more. This will effectively eliminate the need to compute $\phi^{c}\left(f^{\mathcal{G},w^{\prime}}\right)$,
which is responsible for the intractability of the algorithm (\ref{eq:gen-filt-eval-exhaustive}).
The resulting theorem will be a generalization of theorem (\ref{eq:DP-semiring-fusion}).

To apply this idea, we will need constraints $c$ expressed in a \emph{separable
}form, which we explain in detail next. Although not entirely general,
many kinds of constraints typically encountered in combinatorial problems
are in this form \citep{Sniedovich-2011}. Such separable constraints
are formalized using a \emph{constraint algebra} which we denote by
$\mathcal{M}=\left(\mathbb{M},\odot,i_{\odot}\right)$. The only restrictions
on the constraint set $\mathbb{M}$ is that it is countable and (for
practical implementation purposes) of finite cardinality. The binary
operator $\odot$ is usually accompanied by an identity, $i_{\odot}$
(but this is not essential, we will illustrate this below). Then,
a typical separable constraint can be defined in terms of a recurrence
function $h^{\mathcal{M},v}:\mathbb{N}\to\mathbb{M}$ partially applied
to monoid $\mathcal{M}$'s operator $\odot$ and identity $i_{\odot}$
over a set of configurations of length $L$,

\begin{equation}
\begin{aligned}h_{0}^{\mathcal{M},v} & =i_{\odot}\\
h_{l}^{\mathcal{M},v} & =h_{l-1}^{\mathcal{M},v}\odot v\left(x_{l}\right)\quad\forall l\in\left\{ 1,2,\ldots,L\right\} ,
\end{aligned}
\label{eq:constrain_recur}
\end{equation}
where the \emph{constraint map} $v:\mathbb{X}\to\mathbb{M}$ maps
an element $x$ in a configuration $l\in\mathbb{L}$ into a constraint
value $\mathbb{M}$. Widely encountered algebras include arbitrary
\emph{finite monoids }($\odot$ is associative) and \emph{finite groups}.
To complete this separable specification of the constraint, we define
a Boolean \emph{acceptance condition}, $a:\mathbb{M}\to\mathbb{B}$,
whereby a configuration is retained if $a\left(h_{L}^{\mathcal{M},v}\right)$
evaluates to true. Thus, for a configuration $l:\left\{ \left[\mathbb{X}\right]\right\} $
the constraint in (\ref{eq:constrained-semiring-spec}) is computed
using $c\left(l\right)=a\left(h_{L}^{\mathcal{M},v}\right)$.

In this separable formulation, problem (\ref{eq:gen-filt-eval-exhaustive})
is restated by

\begin{equation}
s_{\mathrm{cons}}^{*}=g^{\mathcal{S},w}\left(\phi^{\mathcal{M},v,a}\left(f^{\mathcal{G},w^{\prime}}\right)\right),\label{eq:gen-filt-eval}
\end{equation}
with the modified filtering function $\phi^{\mathcal{M},v,a}:\left\{ \left[\mathbb{X}\right]\right\} \to\left\{ \left[\mathbb{X}\right]\right\} $
which is partially applied to the algebra $\mathcal{M}$, constraint
mapping $v$ and acceptance criteria $a$, implementing $c$. To illustrate,
a specific, recursive implementation of $\phi$ can be given as \citep{bird-1996},
\begin{equation}
\begin{aligned}\phi\left(\emptyset\right) & =\emptyset\\
\phi\left(\left\{ l\right\} \right) & =\begin{cases}
\left\{ l\right\}  & a\left(h_{l}\right)=T\\
\emptyset & \textrm{otherwise}
\end{cases}\\
\phi\left(l_{1}\cup l_{2}\right) & =\phi\left(l_{1}\right)\cup\phi\left(l_{2}\right)
\end{aligned}
\end{equation}
where $l_{1}$, $l_{2}$ are configurations in the configuration set
$\mathbb{L}$ and we suppress the superscripted parameters of $\phi$
and $h$ only for clarity.

To give a concrete, practical example of this separable constraint
formalism, with the additive constraint group $\mathcal{M}=\left(\mathbb{N},+,0\right)$,
the constraint with the mapping $v\left(x\right)=1$ computes lengths
of configurations $l\in\mathbb{L}$ . Indeed, this algebra is just
the \emph{list length }homomorphism defined by the recursion $h_{0}=0$,
$h_{l}=h_{l-1}+1$ \citep{bird-1996}. Thus, the recurrence (\ref{eq:subs-poly})
coupled with this constraint group and the acceptance criteria,

\begin{equation}
a\left(m\right)=\begin{cases}
T & m=M\\
F & \textrm{otherwise}
\end{cases}
\end{equation}
restricts configurations to only those of length $M\in\mathbb{N}$\emph{.}

With this separable formulation, we can construct a new semiring by
\emph{lifting }$\mathcal{S}$ over the algebra $\mathcal{M}$ \citep{Jeuring-1993,emoto-2012}.
This is the desired semiring with which we can fuse the constraint
$c$ with the semiring homomorphism (\ref{eq:semiring-hom}). Lifting
creates vectors (which we can also consider as functions) of semiring
values of type $\mathbb{M}\to\mathbb{S}$. The new, composite semiring
$\mathcal{S}\mathcal{M}=\left(\mathbb{M}\text{\ensuremath{\to\mathbb{S}}},\oplus_{\mathcal{M}},\otimes_{\mathcal{M}},i_{\oplus_{\mathcal{M}}},i_{\otimes_{\mathcal{M}}}\right)$
has binary operators over values $x,y:\mathbb{M}\text{\ensuremath{\to\mathbb{S}}}$given
explicitly by,

\begin{equation}
\begin{aligned}\left(x\oplus_{\mathcal{M}}y\right)_{m} & =x_{m}\oplus y_{m}\\
\left(x\otimes_{\mathcal{M}}y\right)_{m} & =\bigoplus_{\substack{m^{\prime}\odot m^{\prime\prime}=m\\
\forall m^{\prime},m^{\prime\prime}\in\mathbb{M}
}
}\left(x_{m^{\prime}}\otimes y_{m^{\prime\prime}}\right),
\end{aligned}
\label{eq:lifted_semiring-ops}
\end{equation}
where $x_{m}$ refers to indexing with values in $\mathbb{M}$. The
associated operator identities are,

\begin{equation}
\begin{aligned}\left(i_{\oplus_{\mathcal{M}}}\right)_{m} & =i_{\oplus}\quad\forall m\in\mathbb{M}\\
\left(i_{\otimes_{\mathcal{M}}}\right)_{m} & =\begin{cases}
i_{\otimes} & m=i_{\odot}\\
i_{\oplus} & \textrm{otherwise}.
\end{cases}
\end{aligned}
\label{eq:lifted_semiring-ids}
\end{equation}

The operator $\otimes_{\mathcal{M}}$ is an interesting and far-reaching
generalization of discrete convolution operators found in contexts
such as digital signal processing, machine learning and applied mathematics.
We also need the lifted value mapping, $w_{\mathcal{M}}:\mathbb{X}\to\left(\mathbb{M}\to\mathbb{S}\right)$
given by,
\begin{equation}
\left(w_{\mathcal{M}}\left(x\right)\right)_{m}=\begin{cases}
w\left(x\right) & v\left(x\right)=m\\
i_{\oplus} & \textrm{otherwise},
\end{cases}\label{eq:lifted_map}
\end{equation}
where the original semiring map is $w:\mathbb{X}\to\mathbb{S}$ and
the constraint map is $v:\mathbb{X}\to\mathbb{M}$.

Finally, to obtain the solution to $s_{\mathrm{cons}}^{*}$ to (\ref{eq:DP-semiring-fusion}),
we need to \emph{project }the vector lifted over $\mathbb{M}$, onto
$\mathbb{B}$ using the projection function $\pi^{\mathcal{S},a}:\left(\mathbb{M}\to\mathbb{S}\right)\to\mathbb{S}$
partially applied to semiring $\mathcal{S}$ operators and identities
and acceptance criteria $a$,

\begin{equation}
\pi^{\mathcal{S},a}\left(x\right)=\bigoplus_{m^{\prime}\in\mathbb{M}:a\left(m^{\prime}\right)}x_{m^{\prime}}.\label{eq:lifted_accept}
\end{equation}

This yields all the ingredients to define a theorem which we call\emph{
semiring constrained fusion},

\begin{equation}
s_{\mathrm{cons}}^{*}=g^{\mathcal{S},w}\left(\phi^{\mathcal{M},v,a}\left(f^{\mathcal{G},w^{\prime}}\right)\right)=\pi^{\mathcal{S},a}\left(f^{\mathcal{S}\mathcal{M},w_{\mathcal{M}}}\right).\label{eq:DP-filter-fusion}
\end{equation}

To summarize (\ref{eq:DP-filter-fusion}), on the left hand side,
$f^{\mathcal{G},w^{\prime}}$ exhaustively computes configurations
in $\mathbb{L}$; $\phi^{\mathcal{M},v,a}$ filters away any configurations
which do not satisfy the constraint $c$ implemented by $v$ and $a$
over the constraint algebra $\mathcal{M}$ and $g^{\mathcal{S},w}$
evaluates the remaining configurations in $\mathbb{L}$ with the semiring
$\mathcal{S}$ using the mapping $w$. On the right hand side, $f^{S\mathcal{M},w_{\mathcal{M}}}:\left(\mathbb{M}\to\mathbb{S}\right)$
(fully) applied to lifted semiring's $\mathcal{SM}$'s operators and
identities and lifted mapping function $w_{\mathcal{M}}$, first maps
the element $\mathbb{X}$ into lifted semiring values using $w$ and
lifts them over the constraint set using $v$. It then evaluates all
configurations in $\mathbb{L}$, \emph{for every value of the constraint
set }$\mathbb{M}$, using the lifted semiring $\mathcal{S}\mathcal{M}$.
Finally, $\pi^{\mathcal{S},a}:\left(\mathbb{M}\to\mathbb{S}\right)\to\mathbb{S}$
projects this lifted computation back down onto the plain semiring
$\mathcal{S}$.

The proof of (\ref{eq:DP-filter-fusion}) result is given in \nameref{sec:AppendixB}.
To give some informal intuition into the steps in this proof, we show
that (\ref{eq:lifted_semiring-ops}) is a semiring and then using
a change of variables argument, show how the projection is obtained.
We then demonstrate that the composition of the filtering with the
semiring homomorphism, can be fused into a new homomorphism over the
constraint-lifted semiring. This allows us to use our already established
result (\ref{eq:DP-filter-fusion}) to show that the composition of
the exhaustive configuration generator followed by the filtering,
is equivalent to evaluation of these configurations in the lifted
semiring, followed by a projection.

We raise a couple of important comments about this theorem here:
\begin{enumerate}
\item The effect on the computational and memory complexity of the corresponding
unconstrained algorithm (\ref{eq:DP-semiring-fusion}), is quite predictable.
For most practical computational semirings (e.g. max-plus or sum-product)
the operators $\oplus,\otimes$ will be $O\left(1\right)$. In this
setting, for each value of $m\in\mathbb{M}$, the binary operator
$\oplus_{\mathcal{M}}$ is $O\left(1\right)$, and the operator $\otimes_{\mathcal{M}}$
is $O\left(M^{2}\right)$. Thus, in general, applying a constraint
increases the worst-case computational complexity of an existing polymorphic
semiring generator function $f^{\mathcal{S},w}$ by a multiplicative
factor $O\left(M^{3}\right)$. In terms of memory, lifting requires
storing $M$ values per configuration, therefore the memory complexity
increases multiplicatively by a factor $O\left(M\right)$.
\item Lifting is intimately related to the design of DP algorithms found
in textbooks, in the following way. A widely stated, but intuitive
observation, is that designing practical DP algorithms boils down
to identifying a structural decomposition which makes frequent re-use
of sub-problems \citep{kleinberg-2005}. This design principle is
easy to state, but often quite tricky to apply in practice, as it
can depend upon a serendipitous discovery of the right way to parameterize
the problem. However, implicit to the definition of the constraint
operator $\odot$ for constraints $c\left(l\right)$ which can be
implemented in separable algebraic form, is the relationship that
solutions for different values of the constraint algebra have with
each other. The lifted product in (\ref{eq:lifted_semiring-ops})
combines all solutions at every value of the constraint. However,
for each $m\in\mathbb{M}$, the condition $m^{\prime}\odot m^{\prime\prime}=m$
in the product \emph{partitions }the solutions in a way which determines
how the DP sub-problems should be combined. In other words, this partitioning,
coupled with the pairwise summation, determines the dependency structure
of configurations $l\in\mathbb{L}$. For example, in the case of the
simple natural number addition, $m^{\prime}+m^{\prime\prime}=m$,
to find the sub-problem for the case $m=5$ requires us to combine
all sub-problems $\left(m^{\prime},m^{\prime\prime}\right)=\left(1,4\right)$,
$\left(2,3\right)$, $\left(3,2\right)$ and $\left(4,1\right)$,
together. Interestingly, this also demonstrates that DP decompositions
can be performed in ways that are much more general than the fairly
limited descriptions of combining ``smaller'', self-similar problems.
Indeed, it is useful to think of DP decomposition as arising from
a partitioning of the space of the constraint value under the constraint
operator $\odot$, into subsets of pairs of sub-problems for a given
value of $m\in\mathbb{M}$.
\end{enumerate}

\subsection{Simplifying the constraint algebra}

The main problem with the construction (\ref{eq:DP-filter-fusion})
is that the direct computation of $x\otimes_{\mathcal{M}}y$ is quadratic
in the size of $\mathbb{M}$. This is not a problem for small lifting
sets, but for many practical problems we want to apply constraints
which can take on a potentially large set of values, which makes the
naive application of constraint lifting computationally inefficient.
We also know that it is often possible to come up with hand-crafted
DP algorithms which are more efficient than this. We can, however,
substantially improve on this quadratic dependence by noting that
for many semiring polymorphic configuration generators $f^{\mathcal{S},w}:\mathbb{N}\to\mathbb{S}$
given as DP (Bellman) recursions, we need to compute terms of the
form $u\otimes_{\mathcal{M}}w_{\mathcal{M}}\left(x\right)$ for some
general $u:\mathbb{M}\text{\ensuremath{\to\mathbb{S}}}$. Since the
lifted mapping function $w_{\mathcal{M}}\left(x\right)_{m}\ne i_{\oplus}$
only for one value, $m^{\prime\prime}=v\left(x\right)$, we can simplify
the double summation to a single one,
\begin{equation}
\begin{aligned}\left(u\otimes_{\mathcal{M}}w_{\mathcal{M}}\left(x\right)\right)_{m} & =\bigoplus_{\substack{m^{\prime}\odot m^{\prime\prime}=m\\
\forall m^{\prime},m^{\prime\prime}\in\mathbb{M}
}
}\left(u_{m^{\prime}}\otimes w_{\mathcal{M}}\left(x\right)_{m^{\prime\prime}}\right)\\
 & =\left(\bigoplus_{\substack{m^{\prime}\in\mathbb{M}:m^{\prime}\odot v\left(x\right)=m}
}u_{m^{\prime}}\right)\otimes w\left(x\right).
\end{aligned}
\label{eq:lift-product-map}
\end{equation}

Because the operator $\odot$ does not necessarily have inverses,
solutions $m^{\prime}\in\mathbb{M}$ to the equation $m^{\prime}\odot v\left(x\right)=m$
are not necessarily unique. However, we can flip this around and instead
explicitly compute $m=m^{\prime}\odot v\left(x\right)$ for each $m^{\prime}\in\mathbb{M}$.
This leads to an obvious iterative algorithm,

\begin{equation}
\begin{aligned}z & \mapsfrom i_{\oplus_{\mathcal{M}}}\\
z_{m\odot v\left(x\right)} & \mapsfrom z_{m\odot v\left(x\right)}\oplus\left(u_{m}\otimes w\left(x\right)\right)\quad\forall m\in\mathbb{M},
\end{aligned}
\end{equation}
to obtain $u\otimes_{\mathcal{M}}w_{\mathcal{M}}\left(x\right)=z$
at the end of the iteration; here the arrow $\mapsfrom$ denotes setting
the variable on the left, to the value on the right, possibly overwriting
the previous value of the variable. Thus the product (\ref{eq:lift-product-map})
is an inherently $O\left(M\right)$ operation. As a result, semiring
polymorphic generators modified to produce constrained configurations
will have worst-case multiplicative increase in time complexity of
$O\left(M^{2}\right)$, if the semiring product $\otimes$ appears
in the generator only in terms of the form $u\otimes_{\mathcal{M}}w_{\mathcal{M}}\left(x\right)$.
This is true, for instance of all \emph{algebraic path problems }\citep{huang-2008}
and we will encounter several other examples below.

If, additionally, the algebra $\mathcal{M}$ has \emph{inverses} (for
example, if the algebra is a \emph{group}), on fixing $m$ and $m^{\prime}$,
there is a unique (and often analytical) solution to $m^{\prime}\odot m^{\prime\prime}=m$
which we can write as $m^{\prime\prime}=\left(m^{\prime}\right)^{-1}\odot m$.
This also allows us to simplify the lifted semiring product to the
$O\left(M\right)$ computation,

\begin{equation}
\left(x\otimes_{\mathcal{M}}y\right)_{m}=\bigoplus_{m^{\prime}\in\mathbb{M}}\left(x_{m^{\prime}}\otimes y_{\left(m^{\prime}\right)^{-1}\odot m}\right).\label{eq:lift_product-group}
\end{equation}

Note that we often have finite groups where we are not interested
in defining inverses for all elements, for example where we need $y_{\left(m^{\prime}\right)^{-1}\odot m}$
but $\left(m^{\prime}\right)^{-1}\odot m\notin\mathbb{M}$. In that
case, setting $y_{\left(m^{\prime}\right)^{-1}\odot m}=i_{\oplus}$
suffices to appropriately truncate the above product.

For such group lifting algebras, terms of the form $u\otimes_{\mathcal{M}}w_{\mathcal{M}}\left(x\right)$
simplify even further. We can solve $m^{\prime}\odot v\left(x\right)=m$
uniquely to find $m^{\prime}=v\left(x\right)^{-1}\odot m$, so that
the product (\ref{eq:lift-product-map}) can, in this situation, now
be computed as,
\begin{equation}
\begin{aligned}\left(u\otimes_{\mathcal{M}}w_{\mathcal{M}}\left(x\right)\right)_{m} & =\begin{cases}
i_{\oplus} & m\odot v\left(x\right)^{-1}\notin\mathbb{M}\\
u_{m\odot v\left(x\right)^{-1}}\otimes w\left(x\right) & \textrm{otherwise},
\end{cases}\end{aligned}
\label{eq:lifted_product-group}
\end{equation}
which is an $O\left(1\right)$ time operation. Thus, semiring polymorphic
configuration generators can be modified to produce constrained configurations
with multiplicative time complexity increase of only $O\left(M\right)$,
if constraints are expressible in terms of a group algebra. Some examples
of useful, simplified constraint algebras are listed in \nameref{sec:AppendixD}.

\subsection{Taking stock: worked example\label{subsec:working-theory}}

Let us pause to examine how to apply the preceding theory. Consider
the problem of finding the value of the \emph{minimum sum subsequence
}of a list (a subsequence being a sublist of generally non-consecutive
elements of a list). We can specify this simple, unconstrained combinatorial
optimization problem as,

\begin{equation}
s_{\text{subs}}^{*}=\min_{l\in\mathbb{L}}\underset{x\in l}{\sum}x,\label{eq:subseq-spec}
\end{equation}
where the configuration set $\mathbb{L}$ contains all possible subsequences
of a list, of which there are $2^{N}$ for lists of length $N$. For
instance, the subsequences of the list $\left[-2,1,8\right]$ are
\begin{equation}
\mathbb{L}=\left\{ \left[\,\right],\left[-2\right],\left[1\right],\left[8\right],\left[-2,1\right],\left[-2,8\right],\left[1,8\right],\left[-2,1,8\right]\right\} .
\end{equation}

This is a specification (\ref{eq:semiring-spec}) over the \emph{min-plus
}semiring $\mathcal{R}=\left(\mathbb{R},\min,+,\infty,0\right)$.
For a given a length $N$ list $l:\mathbb{N}\to\mathbb{X}$ indexed
as $l_{n}$, a simple, semiring polymorphic generator is given by
the following DP Bellman recursion,

\begin{equation}
\begin{aligned}f_{0}^{\mathcal{S},w} & =i_{\otimes}\\
f_{n}^{\mathcal{S},w} & =f_{n-1}^{\mathcal{S},w}\otimes\left(i_{\otimes}\oplus w\left(l_{n}\right)\right)\quad\forall n\in\left\{ 1,2,\ldots,N\right\} ,
\end{aligned}
\label{eq:subs-poly}
\end{equation}
where $w:\mathbb{X}\to\mathbb{S}$ embeds elements of some given list
$l$ to $\mathcal{S}$-semiring values. Semiring fusion (\ref{eq:DP-semiring-fusion})
tells us that $s_{\text{subs}}^{*}=f_{N}^{\mathcal{R},id}$ given
by

\begin{equation}
\begin{aligned}f_{0}^{\mathcal{R},id} & =0\\
f_{n}^{\mathcal{R},id} & =f_{n-1}^{\mathcal{R},id}+\min\left(0,l_{n}\right)\quad\forall n\in\left\{ 1,2,\ldots,N\right\} ,
\end{aligned}
\label{eq:alg-min-sum-subs}
\end{equation}
is a correct algorithm for solving problem (\ref{eq:subseq-spec})
exactly in $O\left(N\right)$ time complexity, even though the exhaustive
solution would require generating and evaluating $2^{N}$ subsequences
requiring exponential time.

Now, assume that, rather than subsequences, we are interested in \emph{combinations
}of elements of a list (that is, fixed-size subsequences). We can
define a constrained subsequence problem as

\begin{equation}
s_{\text{combs}}^{*}=\min_{l\in\mathbb{L}:\#\left(l\right)=M}\underset{x\in l}{\sum}x,\label{eq: comb-min-sum-spec}
\end{equation}
where the constraint condition $c\left(l\right)=\left(\#\left(l\right)=M\right)$
receives a subsequence, and evaluates to true if that subsequence
has the given length $M$. For instance, if $M=2$, the configurations
$\mathbb{L}$ constrained by sublist length for list $\left[-2,1,8\right]$
is the set
\begin{equation}
\mathbb{L}^{\prime}=\left\{ l\in\mathbb{L}:\#\left(l\right)=2\right\} =\left\{ \left[-2,1\right],\left[-2,8\right],\left[1,8\right]\right\} .
\end{equation}

This sequence length constraint can be formulated using the lifting
algebra $\mathcal{M}=\left(\left\{ 1,\ldots,M\right\} ,+,0\right)$,
constraint mapping function $v\left(n\right)=1$ and acceptance criteria
$a\left(m\right)=T$ if $m=M$ and $F$ otherwise. Inserting the semiring
$\mathcal{SM}$'s operators and lifted mapping $w_{\mathcal{M}}$
into the semiring polymorphic generator recursion (\ref{eq:subs-poly}),
we obtain,

\begin{equation}
\begin{aligned}f_{0,m} & =\left(i_{\otimes_{\mathcal{M}}}\right)_{m}\\
f_{n,m} & =\left(f_{n-1}\otimes_{\mathcal{M}}\left(i_{\otimes_{\mathcal{M}}}\oplus_{\mathcal{M}}w_{\mathcal{M}}\left(n\right)\right)\right)_{m},
\end{aligned}
\label{eq:combs-poly}
\end{equation}
where we suppress $f_{n,m}$ 's superscripts for clarity. The first
line above simplifies to,
\begin{equation}
f_{0,m}=\begin{cases}
i_{\otimes} & m=0\\
i_{\oplus} & \textrm{otherwise,}
\end{cases}
\end{equation}
and the second line can be simplified as follows,

\begin{equation}
\begin{aligned}f_{n,m} & =\left(f_{n-1}\oplus_{\mathcal{M}}f_{n-1}\otimes_{\mathcal{M}}w_{\mathcal{M}}\left(n\right)\right)_{m}\\
 & =f_{n-1,m}\oplus\left(f_{n-1}\otimes_{\mathcal{M}}w_{\mathcal{M}}\left(n\right)\right)_{m}\\
 & =f_{n-1,m}\oplus\begin{cases}
i_{\oplus} & m-1\notin\mathbb{M}\\
f_{n-1,m-1}\otimes w\left(n\right) & \textrm{otherwise}
\end{cases}\\
 & =\begin{cases}
f_{n-1,0} & m=0\\
f_{n-1,m}\oplus\left(f_{n-1,m-1}\otimes w\left(n\right)\right) & \textrm{otherwise},
\end{cases}
\end{aligned}
\label{eq:combs-simple-poly}
\end{equation}
for all $n\in\left\{ 1,2,\ldots,N\right\} $ and $m\in\left\{ 1,2,\ldots,M\right\} $.
Explicitly, by invoking constrained semiring fusion (\ref{eq:DP-filter-fusion}),
in the min-plus semiring $\mathcal{R}$ with $w=id$ this is,
\begin{equation}
\begin{aligned}f_{0,m}^{\mathcal{R},id} & =\begin{cases}
0 & m=0\\
\infty & \textrm{otherwise}
\end{cases}\\
f_{n,m}^{\mathcal{R},id} & =\begin{cases}
f_{n-1,0}^{\mathcal{R},id} & m=0\\
\min\left(f_{n-1,m}^{\mathcal{R},id},f_{n-1,m-1}^{\mathcal{R},id}+l_{n}\right) & \textrm{otherwise}.
\end{cases}
\end{aligned}
\label{eq:combs-min-plus}
\end{equation}
which is a simple $O\left(N\,M\right)$, provably correct algorithm
for solving \ref{eq: comb-min-sum-spec} through computing $s_{\text{combs}}^{*}=\pi^{\mathcal{R},a}\left(f_{N}^{R,id}\right)=f_{N,M}^{\mathcal{R},id}$
using the recursion (\ref{eq:combs-min-plus}) above.

It is instructive to compare this systematically derived algorithm
to the textbook presentation of similar DP algorithms such as the
quasi-polynomial knapsack problem \citep{kleinberg-2005,emoto-2012}.
We have obtained this DP algorithm by starting from a specification
and by provably correct derivation steps, arrived at the new, computationally
efficient recurrence above which solves the constrained problem. Often,
the solutions obtained this way resemble hand-coded DP algorithms
which involve ad-hoc and specific reasoning where we have to resort
to special case analysis to demonstrate correctness and computational
complexity, after the algorithm is coded.

\subsection{Tupling semirings to avoid backtracing\label{subsec:Tupling-semirings}}

The above cases have demonstrated the use of arbitrary semirings where
some scalar-valued, numerical solution is required. It is often the
case for \emph{optimization }problems (involving the use of \emph{selection
semirings }such min-plus, $\mathcal{R}$) that we also want to know
\emph{which }solutions lead to the optimal (semiring) value. The usual
solution to this (given in most DP literature) is \emph{backtracing},
which retains a list of decisions at each stage and a series of ``back
pointers'' to the previous decision, and then recovers the unknown
decisions by following the sequence of pointers backwards.

In fact, there is an alternative and conceptually simpler solution
made possible with the use of appropriate semirings. In particular
we will focus on the generator semiring $\mathcal{G}$. We can always
exploit what is known as the \emph{tupling trick }to compute two different
semirings simultaneously \citep{bird-1996}. If we map the semiring
values used during the DP computations inside a pair $\left(\mathbb{S},\left\{ \left[\mathbb{S}\right]\right\} \right)$,
then we can simultaneously update a semiring total while retaining
the values selected in that stage. For example, the \emph{arg-max-plus}
selection, also known as the \emph{Viterbi}, semiring \citep{goodman-1999,emoto-2012},
\begin{equation}
\mathcal{S}\times\mathcal{G}=\left(\mathbb{S}\times\left\{ \left[\mathbb{S}\right]\right\} ,\oplus,\otimes,\left(-\infty,\emptyset\right),\left(0,\left\{ \left[\,\right]\right\} \right)\right),
\end{equation}
has operators given explicitly by,
\begin{equation}
\begin{aligned}\left(u,x\right)\oplus\left(v,y\right) & =\begin{cases}
\left(u,x\right) & u>v\\
\left(v,y\right) & u<v\\
\left(u,x\cup y\right) & \textrm{otherwise}
\end{cases}\\
\left(u,x\right)\otimes\left(v,y\right) & =\left(u+v,x\circ y\right),
\end{aligned}
\label{eq:semiring_tupling}
\end{equation}
with identities $i_{\oplus}=\left(-\infty,\emptyset\right)$ and $i_{\otimes}=\left(0,\left\{ \left[\,\right]\right\} \right)$.
Furthermore, it is straightforward to construct a semiring which extends
the Viterbi semiring by maintaining a ranked list of optima, i.e.
computing the top $k$\emph{ }optimal solutions, not merely the single
highest scoring one \citep{goodman-1999}.

In most practical situations, for instance, where the mapping function
in the DP problem is real-valued $w:\mathbb{X}\to\mathbb{R}$ and
thus have effectively zero probability of not being unique, there
will only be a single, rather than potentially multiple, optimal solutions.
In that case, we can remove the ambiguities in the selection with
a simpler semiring,

\begin{equation}
\mathcal{S}\times\mathcal{G}=\left(\mathbb{S}\times\left[\mathbb{S}\right],\oplus,\otimes,\left(-\infty,\emptyset\right),\left(0,\left[\,\right]\right)\right),
\end{equation}
with operators,

\begin{equation}
\begin{aligned}\left(u,x\right)\oplus\left(v,y\right) & =\begin{cases}
\left(u,x\right) & u\ge v\\
\left(u,y\right) & u<v
\end{cases}\\
\left(u,x\right)\otimes\left(v,y\right) & =\left(u+v,x\cup y\right).
\end{aligned}
\label{eq:semiring_tupling_simple}
\end{equation}

Clearly, the semiring $\mathcal{S}\times\mathcal{G}$ is the tupling
of max-plus with $\mathcal{G}$ in such a way as to compute both the
value of the optimal solution alongside the values used to compute
it.

Backtracing and the simple (Viterbi) tupled semiring will usually
be similar in terms of computational complexity. Any particular, practical
implementation may well require a more detailed investigation of the
specifics of the particular data structures in the DP recurrence and
the software and/or hardware platforms involved. For instance, while
semiring tupling does require list concatenation and array structures
could certainly pose a complexity issue, when implemented instead
as linked lists this concatenation takes $O\left(1\right)$ time,
and indeed in practice, DP algorithm recurrences derived using the
methods described here reduce to a sequence of list append operations,
$O\left(1\right)$ each for linked lists.

However, the specifics of lower-level implementation complexity are
incidental: the overall argument is one of high-level conceptual clarity
and systematic derivation, since the tupling construction requires
no modification to the algebraic derivations given here. For instance,
backtracing requires a way to traverse the DP decisions correctly
in the reverse order, which will be special to each DP problem. With
tupled semirings and semiring polymorphic generator functions, all
that is required is to change the semiring of the DP generator algorithm
function as described above, we do not need to know anything about
how the generator algorithm works. Additionally, tupling semirings
is trivially compatible with any set of multiple lifting constraints
where hand-coding of special implementations of backtracking through
the same set of multiple interacting constraints, would be error-prone.

\subsection{Constructing new DP algorithms from old}

The above sections have explained how to specify a combinatorial problem
in an arbitrary semiring and how to derive an efficient algorithm
to solve it. It has demonstrated how the use of semiring lifting allows
us to modify an existing DP algorithm specification with a constraint,
which can then be solved efficiently using constrained semiring fusion
(\ref{eq:DP-filter-fusion}). Here, we show how this gives us a way
of creating new, semiring polymorphic DP generators from existing
ones such as $f^{\mathcal{S},w}$. To see this, note that the semiring
homomorphism $g^{\mathcal{G},w^{\prime}}$ with $w^{\prime}\left(x\right)=\left\{ \left[x\right]\right\} $
is the \emph{identity }homomorphism $id^{\mathcal{G}}:\left\{ \left[\mathbb{X}\right]\right\} \to\left\{ \left[\mathbb{X}\right]\right\} $
for values in the semiring $\mathcal{G}$, i.e. it maps sets of lists,
into the same sets of lists unchanged. Thus we obtain,

\begin{equation}
\begin{aligned}\phi^{\mathcal{M},v,a}\left(f^{\mathcal{G},w^{\prime}}\right) & =id^{\mathcal{G}}\left(\phi^{\mathcal{M},v,a}\left(f^{\mathcal{G},w^{\prime}}\right)\right)\\
 & =g^{\mathcal{G},w^{\prime}}\left(\phi^{\mathcal{M},v,a}\left(f^{\mathcal{G},w^{\prime}}\right)\right)\\
 & =\pi^{\mathcal{G},a}\left(f^{\mathcal{G}\mathcal{M},w_{\mathcal{M}}^{\prime}}\right)
\end{aligned}
\end{equation}
where the last step invokes (\ref{eq:DP-filter-fusion}). We have
shown that $\phi^{\mathcal{M},v,a}\left(f^{\mathcal{G},w^{\prime}}\right)=\pi^{\mathcal{G},a}\left(f^{\mathcal{G}\mathcal{M},w_{\mathcal{M}}^{\prime}}\right)$,
which we denote by $f^{\prime\mathcal{G},w^{\prime}}=\phi^{\mathcal{M},v,a}\left(f^{\mathcal{G},w^{\prime}}\right)$,
this is the generator which computes a constrained configuration set.
Now, fixing $\mathcal{M}$, acceptance criteria $a$ and constraint
map $v$, we notice that $f^{\prime\mathcal{G},w^{\prime}}$ depends
only upon the semiring $\mathcal{G}$ and map $w^{\prime}$. Thus
it, too, is semiring polymorphic since we can replace the $\mathcal{G}$
and $w^{\prime}$ in $f^{\prime}$ with \emph{any arbitrary }semiring
$\mathcal{S}$ and mapping $w$ to obtain $f^{\prime\mathcal{S},w}$.
It follows that we can consider $f^{\prime\mathcal{S},w}$ as a new
semiring polymorphic generator function derived from the existing
generator, $f^{\mathcal{S},w}$. This implies, in particular:
\begin{enumerate}
\item If $f^{\mathcal{S},w}$ is the generator of $\mathbb{L}$ for the
problem $s^{*}=\bigoplus_{l\in\mathbb{L}}\bigotimes_{x\in l}w\left(x\right)$
in (\ref{eq:semiring-spec}), then the new, derived polymorphic generator
$f^{\prime\mathcal{S},w}$ is the generator for the \emph{constraint
augmented }problem $s^{\prime*}=\bigoplus_{l\in\mathbb{L}^{\prime}}\bigotimes_{x\in l}w\left(x\right)$
where $\mathbb{L}^{\prime}=\left\{ l\in\mathbb{L}:c\left(l\right)\right\} $
and $c$ is implemented by $\mathcal{M}$, $v$ and $a$ which are
fixed in (implicit to) $f^{\prime\mathcal{S},w}$. This, of course,
just a way of expressing an algorithm for efficiently computing $s_{\mathrm{cons}}^{*}$
in terms of an algorithm for efficiently computing $s^{*}$ which
is implicit to the constrained semiring fusion theorem (\ref{eq:DP-filter-fusion}).
\item Being semiring polymorphic, this new generator function $f^{\prime\mathcal{S},w}$
satisfies all the conditions of semiring fusion (\ref{eq:DP-semiring-fusion}),
i.e. with an arbitrary semiring homomorphism $g^{\mathcal{S},w}$,
we have $g^{\mathcal{S},w}\left(f^{\prime\mathcal{G},w^{\prime}}\right)=f^{\prime\mathcal{S},w}$.
\item We can repeat this process of augmenting an existing specification
solved by use of a semiring polymorphic generator to derive novel,
polymorphic DP algorithm with \emph{multiple}, simultaneous constraints.
This is possible, essentially, because lifting can always be ``\emph{nested}'',
i.e. lifted semirings can themselves be lifted.
\end{enumerate}
These implications have practical consequences in applications, which
we now illustrate. Above, we demonstrated how to derive an algorithm
for the min-plus problem over subsequence combinations (\ref{eq:combs-min-plus})
from the min-plus problem over subsequences, by semiring lifting applied
to the polymorphic generator for subsequences, (\ref{eq:subs-poly})
which we repeat here for convenience,

\begin{equation}
\begin{aligned}f_{0}^{\mathcal{S},w} & =i_{\otimes}\\
f_{n}^{\mathcal{S},w} & =f_{n-1}^{\mathcal{S},w}\otimes\left(i_{\otimes}\oplus w\left(n\right)\right)\quad\forall n\in\left\{ 1,2,\ldots,N\right\} ,
\end{aligned}
\end{equation}
which is known as an efficient DP algorithm for solving the specification
$s_{\text{subspoly}}^{*}=\bigoplus_{l\in\mathbb{L}}\bigotimes_{x\in l}w\left(x\right)$
over an arbitrary semiring $\mathcal{S}$ and mapping $w$ by computing
$s_{\text{subspoly}}^{*}=f_{N}^{\mathcal{S},w}$. The resulting polymorphic
generator of combinations, (\ref{eq:combs-poly}), is a straightforward
DP Bellman recursion for solving specifications over subsequence combinations
of length $M$. For semirings wherein the operators can be evaluated
in constant time, this has $O\left(N\,M\right)$ time complexity.
It is interesting and useful in its own right so we provide a pseudocode
implementation here, Algorithm \ref{alg:DP-combs}.

\begin{algorithm}
\begin{lyxcode}
function~polycombs($\oplus,\otimes,i_{\oplus},i_{\otimes},w,N,M$)

$f\left[0,0\right]=i_{\otimes}$

$f\left[0,1\ldots M\right]=i_{\oplus}$

for~$n=1\ldots N$~

\begin{lyxcode}
for~$m=0\ldots M$

if~$m=0$~

\begin{lyxcode}
$f\left[n,m\right]=f\left[n-1,0\right]$
\end{lyxcode}
else~

\begin{lyxcode}
$f\left[n,m\right]=f\left[n-1,m\right]\oplus f\left[n-1,m-1\right]\otimes w\left(n\right)$
\end{lyxcode}
\end{lyxcode}
return~$f\left[N,M\right]$
\end{lyxcode}
\caption{Procedural pseudocode implementation of a semiring polymorphic DP
Bellman recursion for subsequence combinations, derived systematically
from a polymorphic subsequence recurrence using constraint lifting
and algebraic simplifications described in the text.\label{alg:DP-combs}}
\end{algorithm}

As a somewhat more complex example, for some applications, there is
a need to perform computations over \emph{non-empty subsequences},
that is subsequences without the empty sub-sequence $\left\{ \emptyset\right\} $.
Analogous to the subsequence problem, we can define the non-empty
subsequence problem in an arbitrary semiring as,
\begin{equation}
s_{\text{subsnepoly}}^{*}=\bigoplus_{l\in\mathbb{L}:l\ne\emptyset}\underset{x\in l}{\bigotimes}w\left(x\right),
\end{equation}
where the constraint function $c\left(l\right)=\left(l\ne\emptyset\right)$
is true if the configuration is non-empty. For instance, for the sequence
$\left[1,2,3\right]$, the set of non-empty subsequences consists
of the set 
\begin{equation}
\mathbb{L}^{\prime}=\left\{ l\in\mathbb{L}:l\ne\emptyset\right\} =\left\{ \left[1\right],\left[2\right],\left[3\right],\left[1,2\right],\left[1,3\right],\left[2,3\right],\left[1,2,3\right]\right\} 
\end{equation}
which has size $2^{N}-1$. To use semiring lifting we need a constraint
algebra for $c\left(l\right)$, for which we define an \emph{existence}
constraint algebra $\mathcal{M}=\left(\mathbb{B},\vee,F\right)$ (see
\nameref{sec:AppendixB}) with the constant constraint map $v\left(n\right)=T$
which partitions the set of subsequences generated by the recurrence
(\ref{eq:subs-poly}), into empty $m=F$ and non-empty $m=T$ subsequences,
\begin{equation}
\begin{aligned}f_{0,m} & =\begin{cases}
i_{\otimes} & m=F\\
i_{\oplus} & m=T
\end{cases}\\
f_{n,m} & =\left(f_{n-1}\oplus_{\mathcal{M}}f_{n-1}\otimes w_{\mathcal{M}}\left(n\right)\right)_{m}\\
 & =f_{n-1}\oplus\begin{cases}
w\left(n\right)\otimes\left(f_{n-1,F}\oplus f_{n-1,T}\right) & m=T\\
i_{\oplus} & m=F.
\end{cases}
\end{aligned}
\end{equation}
suppressing the superscript dependence of $f$ here on $\mathcal{S},w$
for clarity. The last line can be rewritten,
\begin{equation}
\begin{aligned}f_{n,m} & =\begin{cases}
f_{n-1,T}\oplus w\left(n\right)\otimes\left(f_{n-1,F}\oplus f_{n-1,T}\right) & m=T\\
f_{n-1,F} & m=F
\end{cases}\\
 & =\begin{cases}
f_{n-1,T}\oplus\left(w\left(n\right)\otimes f_{n-1,F}\right)\oplus\left(w\left(n\right)\otimes f_{n-1,T}\right) & m=T\\
i_{\otimes} & m=F,
\end{cases}
\end{aligned}
\end{equation}
so that $f_{N,F}=i_{\otimes}$ as expected in the empty subsequence
case. Focusing on the case we want, $f_{N,T}$, we have,
\begin{equation}
\begin{aligned}f_{n,T} & =f_{n-1,T}\oplus\left(w\left(n\right)\otimes f_{n-1,T}\right)\oplus w\left(n\right)\end{aligned}
,
\end{equation}
which, being expressed entirely in terms of the case $m=T$, allows
us to ignore the lifting altogether to obtain the following semiring
polymorphic generator for non-empty subsequences,

\begin{equation}
\begin{aligned}f_{0} & =i_{\oplus}\\
f_{n} & =f_{n-1}\oplus\left(f_{n-1}\otimes w\left(n\right)\right)\oplus w\left(n\right)\quad\forall n\in\left\{ 1,2,\ldots,N\right\} .
\end{aligned}
\label{eq:ne-subs-poly}
\end{equation}
which solves $s_{\text{subsnepoly}}^{*}=f_{N}$ and requires only
$O\left(N\right)$ steps.

We now build on this result further by augmenting this recurrence
with additional constraints, to provide a novel class of algorithms
for special kinds of non-empty subsequences. Algorithms derived in
this subsection include solutions to the \emph{longest increasing
subsequence }problem, which occurs frequently in applications such
as computational genomics \citep{zhang-2003}. This problem can be
specified as,
\begin{equation}
s_{\text{lis}}^{*}=\max_{l\in\mathbb{L}:\left(l\ne\emptyset\right)\land\textrm{inc}\left(l\right)}\#\left(l\right),
\end{equation}
where the constraint constraint $\textrm{inc}\left(l\right)$ returns
true if $l$ is an increasing subsequence. This problem is in the
form of (\ref{eq:semiring-spec}) with semiring $\mathcal{S}=\left(\mathbb{N},\max,+,0,0\right)$
with map $w\left(x\right)=1$ which counts the length of the list,
$l$. For example, for the sequence$\left[1,5,2\right]$, the constrained
configuration set is
\begin{equation}
\mathbb{L}^{\prime\prime}=\left\{ l\in\mathbb{L}:\left(l\ne\emptyset\right)\land\textrm{inc}\left(l\right)\right\} =\left\{ \left[1\right],\left[2\right],\left[5\right],\left[1,2\right],\left[1,5\right]\right\} 
\end{equation}
and $s_{\text{lis}}^{*}=2$.

Starting from the non-empty subsequence recurrence (\ref{eq:ne-subs-poly})
derived above, we can augment this with a constraint that the subsequence
elements must be in an \emph{ordered chain} according to some \emph{binary
relation }which we write $x\mathrm{R}y$. For example the ordering
$x<y$ holds that $x$ must be less than $y$. Here, we require a
somewhat more complex relation in which both sequence and the value
must be ordered, so that we can define a lifting algebra using what
we call a \emph{sequential-value ordering} operator,

\begin{equation}
\left(i,x\right)\preceq\left(j,y\right)=\begin{cases}
\left(j,y\right) & \left(i<j\right)\wedge\left(x<y\right)\\
\left(\infty,\infty\right) & \textrm{otherwise},
\end{cases}\label{eq:ordering-op}
\end{equation}
over tuples $\mathbb{M}=\left(\mathbb{N},\mathbb{R}\right)$, where
$\left(\infty,\infty\right)=z_{\preceq}$ is a special tuple which
act like an \emph{annihilator }or zero element. Operator $\preceq$
is left but not right, associative and it does not have an identity,
so, a lifting algebra $\mathcal{M}=\left(\mathbb{M},\preceq,z_{\preceq}\right)$
using this operator, is not a ``standard'' algebra (such as a monoid,
group or semigroup). The lack of identity means that it cannot be
applied to empty sequences. Nonetheless, the acceptance criteria $a\left(m\right)=T$
if $m\ne z_{\preceq}$ and $T$ otherwise, allows us to filter away
non-empty subsequences which are not in sequentially increasing order,
provided the operator is scanned across the subsequence in left-right
order.

To apply this constraint, we can simplify the lifting algebra using
this ordering operator

\begin{equation}
\left(u\otimes_{\mathcal{M}}w_{\mathcal{M}}\left(n\right)\right)_{m}=\begin{cases}
\left(\oplus_{m^{\prime}:\mathbb{M}:m^{\prime}\preceq m}u_{m^{\prime}}\right)\otimes w\left(n\right) & m=v\left(n\right)\\
i_{\oplus} & \textrm{otherwise}.
\end{cases}
\end{equation}
which, when substituted into (\ref{eq:ne-subs-poly}), gives us
\begin{equation}
\begin{aligned}f_{0,m} & =\left(i_{\oplus_{\mathcal{M}}}\right)_{m}\\
f_{n,m} & =\left(f_{n-1}\oplus_{\mathcal{M}}\left(\begin{cases}
\left(\oplus_{m^{\prime}:\mathbb{M}:m^{\prime}\preceq m}f_{n-1,m^{\prime}}\right)\otimes w\left(n\right) & m=v\left(n\right)\\
i_{\oplus} & \textrm{otherwise}
\end{cases}\right)\oplus w_{\mathcal{M}}\left(n\right)\right)_{m},
\end{aligned}
\end{equation}
for all $n,m\in\left\{ 1,2,\ldots,N\right\} $. The first line simplifies
to $f_{0,m}=i_{\oplus}$, and the second line can be manipulated to
obtain
\begin{equation}
\begin{aligned}f_{0,m} & =i_{\oplus}\\
f_{n,m} & =\begin{cases}
f_{n-1,m}\oplus\left(\oplus_{m^{\prime}:\mathbb{M}:m^{\prime}\preceq m}f_{n-1,m^{\prime}}\right)\otimes w\left(n\right)\oplus w\left(n\right) & m=v\left(n\right)\\
f_{n-1,m} & \textrm{otherwise}.
\end{cases}
\end{aligned}
\label{eq:DP-order-ne-subs-poly}
\end{equation}

To implement this DP recurrence, we next need to choose the lifting
set $\mathbb{M}$. In this setting, we will typically have a unique
(finite) list, e.g. one unique value $u_{n}\in\mathbb{R}$ per $n\in\left\{ 1,2,\ldots,N\right\} $.
Thus, the lifting set consists of the values from this set, e.g. $\mathbb{M}=\left\{ \left(n,u_{n}\right),n\in\left\{ 1,2,\ldots,N\right\} \right\} $,
and the lift mapping functions merely index this set, e.g. $v\left(n\right)=\left(n,u_{n}\right)$.
Note that with this particular lifting set, there is a one-one mapping
between $n$ and any $m:\mathbb{M}$, thus, we can reduce the lifting
set to $\mathbb{M}=\left\{ 1,2,\ldots,N\right\} $ and lift mapping
to $v\left(n\right)=n$, so that the ordering operator becomes,

\begin{equation}
i\preceq j=\begin{cases}
j & \left(i<j\right)\wedge\left(u_{i}<u_{j}\right)\\
\infty & \textrm{otherwise}.
\end{cases}
\end{equation}

These reductions allow us to simplify the above recurrence \ref{eq:DP-order-ne-subs-poly}
to

\begin{equation}
\begin{aligned}f_{0,m} & =i_{\oplus}\\
f_{n,n} & =f_{n-1,n}\oplus\left(\oplus_{m^{\prime}\in\left\{ 1,2,\ldots,n-1\right\} :\left(u_{m^{\prime}}<u_{n}\right)}f_{n-1,m^{\prime}}\right)\otimes w\left(n\right)\oplus w\left(n\right)\\
f_{n,m} & =f_{n-1,m}.
\end{aligned}
\end{equation}

Finally, note that, the second line adds a constant term $w\left(n\right)$
to each $f_{n,n}$, $\oplus$ is associative, and the value of the
first line is independent of $m$, we can move this term from the
second line to the first, leading to the following polymorphic DP
recursion for increasing sequential subsequences,
\begin{equation}
\begin{aligned}f_{0,m} & =w\left(m\right)\\
f_{n,n} & =f_{n-1,n}\oplus\left(\oplus_{m^{\prime}\in\left\{ 1,2,\ldots,n-1\right\} :\left(u_{m^{\prime}}<u_{n}\right)}f_{n-1,m^{\prime}}\right)\otimes w\left(n\right)\\
f_{n,m} & =f_{n-1,m},
\end{aligned}
\label{eq:DP-order-list-ne-subs}
\end{equation}
with the projection $\pi^{\mathcal{S},a}\left(f_{N}\right)=\bigoplus_{m\in\left\{ 1,2,\ldots,N\right\} }f_{N,m}$.
In terms of computational complexity, the recurrence must be computed
for all $n,m\in\left\{ 1,2,\ldots,N\right\} $ and the second requires
$O\left(N\right)$ steps. Note that, the third line does not change
the value of $f_{n,m}$ for $m\ne n$ obtained at the previous iteration,
so that, iterating over $m$, only the term $f_{n,n}$ needs updating
in the second line. Thus, the algorithm requires $O\left(N^{2}\right)$
steps.

The longest increasing sub-sequences DP algorithm which computes $s_{\mathrm{lis}}^{*}$
is obtained as a special case of (\ref{eq:DP-order-list-ne-subs})
with the semiring $\mathcal{S}=\left(\mathbb{N},\max,+,0,0\right)$
and the lift mapping $w\left(n\right)=1$,
\begin{equation}
\begin{aligned}f_{0,m} & =1\\
f_{n,n} & =\max\left(f_{n-1,n},\max_{m^{\prime}\in\left\{ 1,2,\ldots,n-1\right\} :\left(u_{m^{\prime}}<u_{n}\right)}f_{n-1,m^{\prime}}\right)+1\\
f_{n,m} & =f_{n-1,m},
\end{aligned}
\label{eq:DP-alg-lis}
\end{equation}
so that $s_{\mathrm{lis}}^{*}=\pi^{\mathcal{S},a}\left(f_{N}\right)=\max_{m\in\left\{ 1,2,\ldots,N\right\} }f_{N,m}$.
Compared to existing, classical implementations of this algorithm
in the literature \citep{zhang-2003}, we note that, the algebraic
simplifications afforded by our approach makes it transparent that
there is no need to perform $N$ semiring products $\otimes$ in the
second line, which may lead to computational savings in practice.

Whilst, for the longest increasing subsequences problem, there are
somewhat more efficient algorithms which exploit the special structure
of the problem, the generalized ordered subsequences DP algorithm
derived here, (\ref{eq:DP-order-ne-subs-poly}), being polymorphic,
can be applied to any arbitrary binary relation $\mathrm{R}$:
\begin{equation}
x\odot y=\begin{cases}
y & x\mathrm{R}y\\
z_{\odot} & \textrm{otherwise}.
\end{cases}
\end{equation}

For example, we immediately obtain an algorithm for semiring computations
over all non-decreasing subsequences (ordering $x\le y$), or, for
subsequences consisting of sets, all subsequences ordered by inclusion,
$x\subseteq y$.

\section{Applications\label{sec:Applications}}

In this section we will investigate some practical applications of
the algebraic theory developed above.

\subsection{Segmentation}

A problem of perennial importance in statistics and signal processing
is that of \emph{segmentation}, or dividing up a sequence of data
items or a time series $y_{n}$ for $n\in\left\{ 1,2,\ldots,N\right\} $,
into contiguous, non-overlapping intervals (segments) $\left(i,j\right)$
for $i,j\in\left\{ 1,2,\ldots,N\right\} $ with $i\le j$. Thus, $\left(i,i\right)$
is a segment of length one. An example is the problem of (1D) \emph{piecewise
regression}, which involves fitting a functional curve $f\left(n,c_{i,j}\right)$
to segments, and minimizing the sum of model fit errors $E\left(l\right)=\sum_{\left(i,j\right)\in l}e_{i,j}$,
where $e_{i,j}=\frac{1}{p}\sum_{n=i}^{j}\left|y_{n}-f\left(n,c_{i,j}\right)\right|^{p}$
for $p>0$ being the fit error within each interval $\left(i,j\right)$.
The optimal model parameters $c_{i,j}$ can be estimated using any
statistical model-fitting procedure \citep{little-2019}. Thus, the
input of the problem is the time series $y_{n}$, and the output,
after solving the DP problem, will be a set of indicators, from which
a complete model of the time series can be recovered.

Similar to equation (\ref{eq:semiring-spec}), we can specify the
segmentation problem as a combinatorial optimization problem,

\begin{equation}
s_{\text{seg}}^{*}=\min_{l\in\mathbb{L}}E\left(l\right),\label{eq:segment-spec}
\end{equation}
where $\mathbb{L}$ is the all possible segmentation. For $N=4$,
the set of configurations is 
\begin{align}
\mathbb{L}= & \left\{ \left[\left(1,4\right)\right],\left[\left(1,1\right),\left(2,4\right)\right],\left[\left(1,2\right),\left(3,4\right)\right],\left[\left(1,3\right),\left(4,4\right)\right],\right.\nonumber \\
 & \left.\left[\left(1,1\right),\left(2,3\right),\left(4,4\right)\right],\left[\left(1,1\right),\left(2,2\right),\left(3,4\right)\right],\left[\left(1,1\right),\left(2,2\right),\left(3,3\right),\left(4,4\right)\right]\right\} .
\end{align}

An $O\left(N^{2}\right)$ DP algorithm for this problem was devised
by Richard Bellman as follows \citep{kleinberg-2005}. The optimal
segmentation ending at index $j$ can be obtained by combining all
the ``smaller'' optimal segmentations $\left(\ldots,i-1\right)$ with
the following segments $\left(i,j\right)$, for all $i\in\left\{ 1,2,\ldots,j\right\} $.
This gives rise to the following Bellman recursion,

\begin{equation}
\begin{aligned}f_{0} & =0\\
f_{j} & =\min_{i\in\left\{ 1,2,\ldots,j\right\} }\left(f_{i-1}+e_{i,j}\right)\quad\forall j\in\left\{ 1,2,\ldots,N\right\} .
\end{aligned}
\label{eq:DP_segment}
\end{equation}

This recursion efficiently solves the problem (\ref{eq:segment-spec}),
so that $s_{\text{seg}}^{*}=f_{N}$.

Using the theory in Section \ref{sec:Theory}, by \emph{abstracting
}this recursion over an arbitrary semiring $\mathcal{S}$ and semiring
map function $w$, we obtain the polymorphic segmentation generator
algorithm,
\begin{equation}
\begin{aligned}f_{0}^{\mathcal{S},w} & =i_{\otimes}\\
f_{j}^{\mathcal{S},w} & =\bigoplus_{i\in\left\{ 1,2,\ldots,j\right\} }\left(f_{i-1}^{\mathcal{S},w}\otimes w\left(i,j\right)\right)\quad\forall j\in\left\{ 1,2,\ldots,N\right\} .
\end{aligned}
\label{eq:DP_segment-poly}
\end{equation}

Using this polymorphic recursion, we can, for example, obtain the
optimal configuration $l_{\mathrm{seg}}^{*}$ as well $s_{\text{seg}}^{*}=E^{*}$,
using the tupled selection semiring, see Section \ref{subsec:Tupling-semirings}.

Since the segment fit errors are, generally, all non-negative, $e_{i,j}>0$
and shorter segments are typically more accurately modelled than larger
segments (given the same model structure across segments), the problem
as stated above usually has a ``degenerate'' optimal solution with
only the `diagonal' segments $\left(i,i\right),i\in\left\{ 1,2,\ldots,N\right\} $
of length 1, selected. To avoid the collapse onto this degenerate
solution, we can \emph{regularize }the sum \citep{little-2019},

\begin{equation}
s_{\text{segreg}}^{*}=\min_{l\in\mathbb{L}}\left(E\left(l\right)+\lambda\#\left(l\right)\right),
\end{equation}
for the regularization constant $\lambda>0$ where $\#\left(l\right)=\sum_{\left(i,j\right)\in l}1$
counts the number of segments in the configuration $l$. It is simple
to modify the semiring polymorphic DP recursion (\ref{eq:DP_segment-poly})
to include this regularization term simply by choosing the semiring
mapping $w\left(i,j\right)=e_{i,j}+\lambda$, so that $s_{\text{segreg}}^{*}=f_{N}^{\mathcal{R},w}$
using the min-sum semiring $\mathcal{R}$.

\begin{figure}
\begin{centering}
\includegraphics[scale=0.5]{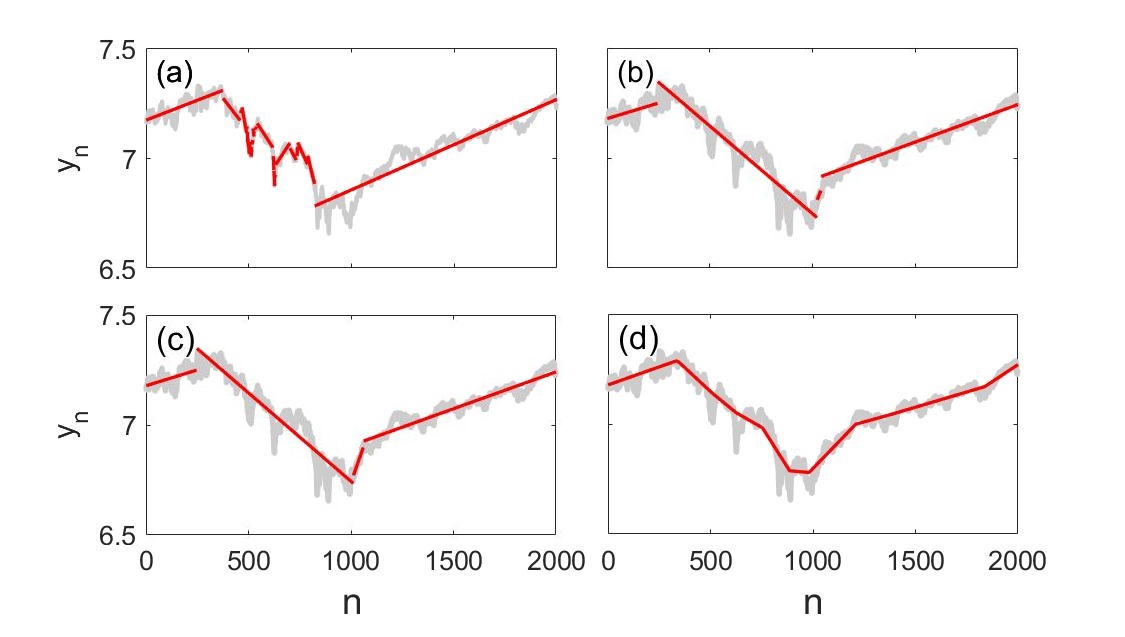}
\par\end{centering}
\caption{DP segmentation algorithms derived using our novel algebraic framework
for solving constrained, 1D segmented, least-squares linear regression,
applied to a sample of logarithmically-transformed S\&P500 financial
index daily values. Input data $y_{n}$ (grey lines), segmentation
result (red line). (a) Unconstrained segmentation with regularization
$\lambda=1.78\times10^{-5}$ , (b) with fixed number segments $L=4$,
(c) with minimum segment length $M=50$ days, and (d) for comparison,
L1 trend filtering with regularization $\lambda=100$.\label{fig:DP_segment_snp500}}
\end{figure}

While this regularization approach is simple, it does not offer much
control over the segmentation quality, as the appropriate choice of
the single parameter $\lambda$ can be difficult to obtain. For example,
some choices lead to over and under-fitting in different parts of
the same signal, see Figure \ref{fig:DP_segment_snp500}(a). Instead,
a more effective level of control can be obtained by directly constraining
the segmentation to a fixed number of segments, which we can express
with the specification, 
\begin{equation}
\begin{aligned}s_{\text{seglen}}^{*} & =\min_{l\in\mathbb{L}:\#\left(l\right)=L}E\left(l\right)\end{aligned}
.\label{eq:seglen-min-sum-spec}
\end{equation}

This is a constrained semiring problem of the form (\ref{eq:constrained-semiring-spec})
where $c\left(l\right)=T$ if the size of $l$ is $L$, over the min-plus
semiring $\mathcal{R}$. To apply semiring lifting, the constraint
algebra needs to count the number of segments up to the fixed number
of segments $L$, which implies we need the lifting algebra $\mathcal{M}=\left(\left\{ 1,2,\ldots,L\right\} ,+,0\right)$
and constraint mapping function $v\left(i,j\right)=1$, with acceptance
condition $a\left(m\right)=T$ if $m=L$. Next, inserting the corresponding
lifted semiring into (\ref{eq:DP_segment-poly}) we obtain,
\begin{equation}
\begin{aligned}f_{0,m}^{\mathcal{S},w} & =\left(i_{\otimes_{\mathcal{M}}}\right)_{m}\\
f_{j,m}^{\mathcal{S},w} & =\left(\underset{i\in\text{\ensuremath{\left\{  1,2,\ldots j\right\} } }}{\oplus_{\mathcal{M}}}\left(f_{i-1}^{\mathcal{S},w}\otimes_{\mathcal{M}}w_{\mathcal{M}}\left(i,j\right)\right)\right)_{m}\quad\forall j\in\left\{ 1,2,\ldots,N\right\} .
\end{aligned}
\label{eq:DP_seglift-poly}
\end{equation}

The first line above simplifies to $f_{0,m}^{\mathcal{S},w}=i_{\otimes}$
for $m=0$ and $i_{\oplus}$ otherwise, and the second line becomes,

\begin{equation}
\begin{aligned}f_{j,m}^{\mathcal{S},w} & =\bigoplus_{i\in\left\{ 1,2,\ldots,j\right\} }\left(f_{i-1}^{\mathcal{S},w}\otimes_{\mathcal{M}}w_{\mathcal{M}}\left(i,j\right)\right)_{m}\\
 & =\bigoplus_{i\in\left\{ 1,2,\ldots,j\right\} }\begin{cases}
i_{\oplus} & m-1\notin\mathbb{M}\\
f_{i-1,m-1}^{\mathcal{S},w}\otimes w\left(i,j\right) & \textrm{otherwise}
\end{cases}\\
 & =\begin{cases}
i_{\oplus} & m=0\\
\bigoplus_{i\in\left\{ 1,2,\ldots,j\right\} }f_{i-1,m-1}^{\mathcal{S},w}\otimes w\left(i,j\right) & \textrm{otherwise},
\end{cases}
\end{aligned}
\label{eq:segs-fixlen-poly}
\end{equation}
using the group product simplification (\ref{eq:lifted_product-group})
in the second step. Finally, to solve (\ref{eq:seglen-min-sum-spec}),
we apply the acceptance condition in the min-sum semiring $\mathcal{R}$,
to obtain
\begin{equation}
s_{\text{seglen}}^{*}=\pi^{\mathcal{R},a}\left(f_{N}^{\mathcal{R},w}\right)=f_{N,L}^{\mathcal{R},w},
\end{equation}
which computes the result in $O\left(N^{2}L\right)$ time with $O\left(N\,L\right)$
memory. In practice, this algorithm produces much more predictable
results than the basic algorithm, see Figure \ref{fig:DP_segment_snp500}(b).
Interestingly, it is well-known in machine learning circles that the
ubiquitous \emph{$K$-means clustering }problem \citep{little-2019},
which is computationally intractable for non-scalar data items and
therefore approximated using heuristic algorithms, can be solved exactly
using the algorithm derived above for scalar data \citep{gronlund-2018}.
However, existing presentations in the literature are not formally
proven correct and thus our presentation here is, to our knowledge,
the first formally correct derivation from specification.

Furthermore, it is trivial to adapt the acceptance criteria $a$ above
to e.g. solve constraints of the form $L^{\prime}\le\#\left(l\right)\le L$,
giving an upper and lower bound on the number of segments, by modifying
$a\left(m\right)=T$ when $L^{\prime}\le m\le L$. The corresponding
DP algorithm is derived from the specification as follows,

\begin{equation}
\begin{aligned}s_{\text{segrange}}^{*} & =\min_{l\in\mathbb{L}:L^{\prime}\le\#\left(l\right)\le L}E\left(l\right)\\
 & =\pi^{\mathcal{R},a}\left(f_{N}^{\mathcal{R},w}\right)\\
 & =\min_{L^{\prime}\le m\le L}f_{N,m}^{\mathcal{R},w}.
\end{aligned}
\end{equation}
where here, $f$ is the semiring polymorphic generator given in (\ref{eq:segs-fixlen-poly}).

The segment count constraint above is fairly straightforward and has
been (re)-invented in an ad-hoc manner before \citep{terzi-2006}.
We will next show how to derive a segmentation algorithm for segmentation
problems with much more elaborate constraints which would be far more
difficult to derive without systematic analytical tools such as we
describe in this paper. While the segment count constraint is certainly
very practical, there are other ways to control the segmentation since
we may not know the number of segments in advance. The \emph{length,
}$d\left(i,j\right)=j-i+1$, of each segment is a property of key
practical importance. For example, it would be extremely useful in
many applications to control the minimum length of each segment,

\begin{equation}
\begin{aligned}s_{\text{segminlen}}^{*} & =\min_{l\in\mathbb{L}:\min_{\left(i,j\right)\in l}d\left(i,j\right)=L}E\left(l\right)\end{aligned}
\end{equation}
which is in the form of (\ref{eq:constrained-semiring-spec}) using
the constraint $c\left(l\right)=T$ if $\left(\min_{x\in l}d\left(x\right)\right)=L$,
i.e. the set of lengths of all the segments in $l$ is at least $L$,
over the min-sum semiring $\mathcal{R}$. Following the procedure
above, we have the lifting algebra $\mathcal{M}=\left(\left\{ 1,2,\ldots,N\right\} ,\min,N\right)$
and lift mapping function $v\left(i,j\right)=j-i+1$. For the semiring
lifted segmentation recursion (\ref{eq:DP_seglift-poly}), the first
line becomes,
\begin{equation}
f_{0,m}=\begin{cases}
i_{\otimes} & m=N\\
i_{\oplus} & \textrm{otherwise}.
\end{cases}
\end{equation}

We also need the product (\ref{eq:lift-product-map}), which becomes,

\begin{equation}
\begin{aligned}\left(u\otimes_{\mathcal{M}}w_{\mathcal{M}}\left(i,j\right)\right)_{m} & =\left(\bigoplus_{\substack{m^{\prime}\in\left\{ 1,2,\ldots,N\right\} \\
\min\left(m^{\prime},d\left(i,j\right)\right)=m
}
}u_{m^{\prime}}\right)\otimes w\left(i,j\right)\end{aligned}
.\label{eq:lift_product-map-segfixlen}
\end{equation}

This lifting algebra is a monoid without analytical (and unique) inverses,
so, to make progress, we need to find an explicit expression for the
set $\left\{ \min\left(m^{\prime},d\left(i,j\right)\right)=m\right\} $
for $m^{\prime}\in\left\{ 1,2,\ldots,N\right\} $. There are three
cases to consider,
\begin{align}
\left\{ m^{\prime}:\min\left(m^{\prime},d\left(i,j\right)\right)=m\right\}  & =\begin{cases}
\left\{ m\right\}  & m<d\left(i,j\right)\\
\left\{ m,m+1,\ldots,N\right\}  & m=d\left(i,j\right),\\
\emptyset & m>d\left(i,j\right)
\end{cases}
\end{align}
and upon inserting this into the product above, we arrive at,
\begin{equation}
\begin{aligned}\left(u\otimes_{\mathcal{M}}w_{\mathcal{M}}\left(i,j\right)\right)_{m} & =\left(\begin{cases}
u_{m} & m<d\left(i,j\right)\\
\oplus_{m^{\prime}=m}^{N}u_{m^{\prime}} & m=d\left(i,j\right)\\
i_{\oplus} & m>d\left(i,j\right)
\end{cases}\right)\otimes w\left(i,j\right)\end{aligned}
\end{equation}
so that the second line of the lifted segmentation recursion (\ref{eq:DP_seglift-poly})
can be simplified,
\begin{equation}
\begin{aligned}f_{j,m}^{\mathcal{S},w} & =\bigoplus_{i\in\left\{ 1,2,\ldots,j\right\} }\left(f_{i-1}\otimes_{\mathcal{M}}w_{\mathcal{M}}\left(i,j\right)\right)_{m}\\
 & =\bigoplus_{i\in\left\{ 1,2,\ldots,j\right\} }\left(\begin{cases}
f_{i-1,m}^{\mathcal{S},w} & m<d\left(i,j\right)\\
\oplus_{m^{\prime}=m}^{N}f_{i-1,m^{\prime}}^{\mathcal{S},w} & m=d\left(i,j\right)\\
i_{\oplus} & m>d\left(i,j\right)
\end{cases}\right)\otimes w\left(i,j\right)\\
 & =\bigoplus_{i\in\left\{ 1,2,\ldots,j\right\} }\begin{cases}
f_{i-1,m}^{\mathcal{S},w}\otimes w\left(i,j\right) & m<d\left(i,j\right)\\
\left(\oplus_{m^{\prime}\in\left\{ m,m+1,\ldots,N\right\} }f_{i-1,m^{\prime}}^{\mathcal{S},w}\right)\otimes w\left(i,j\right) & m=d\left(i,j\right)\\
i_{\oplus} & m>d\left(i,j\right)
\end{cases}
\end{aligned}
\label{eq:DP_segfixlen-poly}
\end{equation}
for all $j\in\left\{ 1,2,\ldots,N\right\} $. Using the acceptance
condition $a\left(m\right)=T$ if $m=L$ we have an $O\left(N^{3}\right)$
time DP algorithm to find the required solution, $s_{\text{segminlen}}^{*}=\pi^{\mathcal{R},a}\left(f_{N}^{\mathcal{R},w}\right)=f_{N,L}^{\mathcal{R},w}$
using the derived recurrence above and min-plus semiring $\mathcal{R}$.
As previously, a simple modification of the acceptance function $a\left(m\right)$
allows, for example, computing optimal segmentations across a range
$L^{\prime}\le\min_{\left(i,j\right)\in l}d\left(i,j\right)\le L$
of minimum segment lengths. Applied to the scalar $K$-means problem,
this modification would be a viable approach to avoiding the problem
of \emph{degenerate clusters} assigned few or no items \citep{little-2019}.

\begin{figure}
\begin{centering}
\includegraphics[scale=0.5]{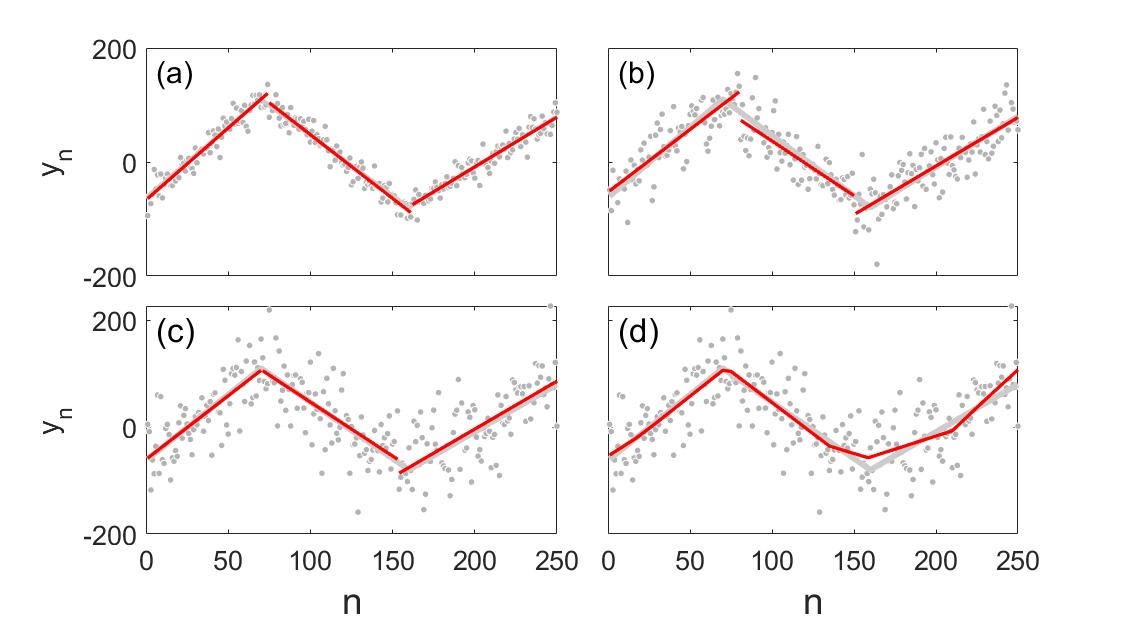}
\par\end{centering}
\caption{DP segmentation algorithms derived using our novel algebraic framework
for solving constrained, 1D segmented, least-squares linear regression,
applied to synthetic, piecewise linear time series with i.i.d. Gaussian
noise, standard deviation $\sigma$. Input data $y_{n}$ (grey dots),
underlying piecewise constant signal (grey line), segmentation result
(red line). (a) Unconstrained segmentation with regularization $\lambda=15$,
noise $\sigma=15$, (b) with fixed number segments $L=3$, noise $\sigma=30$,
(c) with minimum segment length $M=70$, noise $\sigma=60$, and (d)
for comparison, L1 trend filtering with regularization $\lambda=10^{3}$.\label{fig:DP_segment_syn}}
\end{figure}

We find that constrained DP segmented regression, derived using the
algebraic methods introduced here, usually produces very interpretable
results, even for problems where the segmentation boundaries may be
quite difficult to determine using other methods, particularly when
the signal-to-noise ratio is low, see Figure \ref{fig:DP_segment_syn}.
For example, methods such as \emph{L1 trend filtering} \citep{kim-2009}
suffer from the problem that there is often no single, unambiguous
segmentation, see for example Figure \ref{fig:DP_segment_syn}(d)
and Figure \ref{fig:DP_segment_snp500}(d). This is because it is
better to consider such L1-based methods as \emph{smoothing} algorithms
arising from a \emph{convex relaxation} of the combinatorial segmentation
problem. This clearly shows the advantage of constrained, exact combinatorial
optimization in applications such statistical time series analysis,
made practical by the algebraic approach described in this paper.


\subsection{Sequence alignment}

Our next application focus is \emph{sequence alignments}, a problem
of central importance to computational biology, natural language processing
and signal processing. For example, in genomic sequence analysis,
we are often interested in knowing how closely related two DNA or
RNA base pair sequences are, and this can be assessed by computing
the most plausible series of mutations (insertions and deletions)
needed in order to bring the two sequences $u$ and $v$ into alignment.
This alignment problem is specified as an \emph{optimal matching problem
}finding the \emph{minimum cost transformation }from sequence $u$
into sequence $v$,

\begin{equation}
s_{\text{align}}^{*}=\min_{l\in\mathbb{L}}\underset{\left(i,j\right)\in l}{\sum}w\left(i,j\right),\label{eq:seqalign-spec-min-sum}
\end{equation}
where configurations set $\mathbb{L}$ consists of all possible sequence
transformations. These transformations are restricted to insertions,
deletions or simple matches (no transformation required) for pair
sequences at positions $\left(i,j\right)$ in the sequences, $u_{i}$
and $v_{j}$. More specifically, an insert operation aligns the sequences
at position pair $\left(i,j\right)$ with the pair at $\left(i-1,j\right)$
incurring a cost $w\left(0,j\right)$, a deletion aligns the sequences
at position pair $\left(i,j\right)$ with the pair at $\left(i,j-1\right)$
at cost $w\left(i,0\right)$, and a match aligns the sequences at
pair $\left(i,j\right)$ with the pair at $\left(i-1,j-1\right)$
incurring cost $w\left(i,j\right)$. For example, for sequences of
length $\#\left(u\right)=2$ and $\#\left(v\right)=1$, there are
5 possible ways of transforming $u$ into $v$, so that the set of
transformations is,

\begin{equation}
\begin{aligned}\mathbb{L} & =\left\{ \left[\left(1,0\right),\left(2,1\right)\right],\left[\left(1,1\right),\left(0,1\right)\right],\right.\\
 & =\left.\left[\left(0,1\right),\left(0,1\right),\left(0,1\right)\right],\left[\left(1,0\right),\left(1,0\right),\left(0,1\right)\right],\left[\left(1,0\right),\left(2,0\right),\left(2,0\right)\right]\right\} 
\end{aligned}
\end{equation}

One of the earliest and most widely used methods for minimizing this
cost is the \emph{Needleman-Wunsch} (NW) DP algorithm \citep{pachter-2005},
the usual presentation of which is given in the \emph{min-sum }semiring,

\begin{equation}
\begin{aligned}f_{0,0} & =0\\
f_{i,0} & =f_{i-1,0}+w\left(i,0\right)\\
f_{0,j} & =f_{0,j-1}+w\left(0,j\right)\\
f_{i,j} & =\min\left(f_{i-1,j-1}+w\left(i,j\right),f_{i-1,j}+w\left(0,j\right),f_{i,j-1}+w\left(i,0\right)\right)
\end{aligned}
\label{eq:DP_seqalign-minsum}
\end{equation}
for all $i\in\left\{ 1,2,\ldots,N\right\} $ and $j\in\left\{ 1,2,\ldots,M\right\} $,
where $w\left(i,j\right)$ is the cost of the alignment of the first
sequence at position $i$, with the second sequence at position $j$.
The optimal alignment is obtained at $s_{\text{align}}^{*}=f_{N,M}$.

To apply our results, we construct a semiring polymorphic abstraction
of the above,
\begin{equation}
\begin{aligned}f_{0,0}^{\mathcal{S},w} & =i_{\otimes}\\
f_{i,0}^{\mathcal{S},w} & =f_{i-1,0}^{\mathcal{S},w}\otimes w\left(i,0\right)\\
f_{0,j}^{\mathcal{S},w} & =f_{0,j-1}^{\mathcal{S},w}\otimes w\left(0,j\right)\\
f_{i,j}^{\mathcal{S},w} & =\left(f_{i-1,j-1}^{\mathcal{S},w}\otimes w\left(i,j\right)\right)\oplus\left(f_{i-1,j}^{\mathcal{S},w}\otimes w\left(0,j\right)\right)\oplus\left(f_{i,j-1}^{\mathcal{S},w}\otimes w\left(i,0\right)\right)
\end{aligned}
\label{eq:DP_seqalign-poly}
\end{equation}

We can now put this semiring polymorphic generator to use for various
purposes. One application is counting all possible alignments. Semiring
fusion (\ref{eq:DP-semiring-fusion}) tells us that $s_{\mathrm{aligncount}}^{*}=\sum_{l\in\mathbb{L}}\prod_{\left(i,j\right)\in l}1=f_{N,M}^{\mathcal{N},w}$
using the \emph{count semiring} $\mathcal{N}=\left(\mathbb{N},+,\times,0,1\right)$
with $w\left(i,j\right)=1$. A closed-form formula for this number
of alignments, denoted $D\left(N,M\right)$ is not known, but by expanding
$f^{\mathcal{N},w}$ above we find $f_{0,0}=1$, $f_{i,0}=f_{i-1,0}$,
$f_{0,j}=f_{0,j-1}$ and $f_{i,j}=f_{i-1,j-1}+f_{i-1,j}+f_{i,j-1}$,
which simplifies to the following recurrence,
\begin{equation}
\begin{aligned}D\left(n,m\right) & =\begin{cases}
1 & \left(n=0\right)\vee\left(m=0\right)\\
D\left(n-1,m-1\right)+D\left(n-1,m\right)+D\left(n,m-1\right) & \textrm{otherwise}.
\end{cases}\end{aligned}
\label{eq:DP_seqalign-count}
\end{equation}

\begin{figure}
\begin{centering}
\includegraphics[scale=0.08]{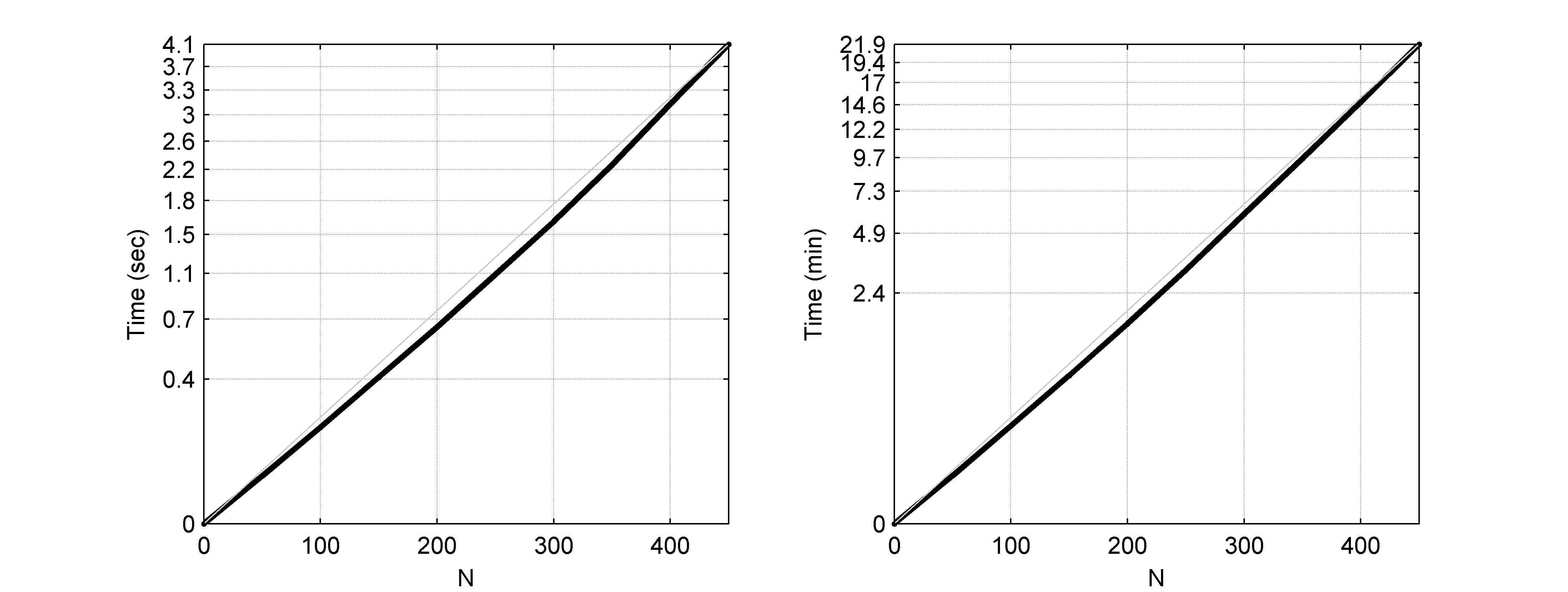}
\par\end{centering}
\caption{Computational time (black line) required to solve the Needleman-Wunsch\emph{
}DP sequence alignment algorithm (left) without constraints and (right)
with lifted constraint. The horizontal axis is the length of both
sequences and also the size of the constraint algebra (e.g. $N=M=\left|\mathcal{M}\right|$).
The vertical axis is on a quadratic (left) and cubic (right) scale
such that exact $O\left(N^{2}\right)$ and $O\left(N^{3}\right)$
complexities correspond to a straight line (grey line). Python language
implementation on a quad-core Intel Core i7 3.2GHz, 16Gb DRAM.\label{fig:align_fwd_timings}}
\end{figure}

This recursion describes the well-known \emph{Delannoy numbers} which
for \emph{$M=N$} is the integer sequence \citet[sequence A001850]{sloane-2021},
with leading order asymptotic approximation $D\left(N,N\right)\approx5.8^{N}$.
Thus, semiring polymorphism allows us to show that brute-force computation
of all alignments would be intractable as it requires exponential
time complexity, whereas the factorized DP NW implementation has $O\left(N\,M\right)$
e.g. quadratic, computational cost (Figure \ref{fig:align_fwd_timings}).

One practical problem with the standard NW algorithm (\ref{eq:DP_seqalign-minsum})
is that it places no constraint on how far the sequences can become
out of alignment. After all, any two DNA/RNA sequences are related
by an arbitrary number of insertions/deletions, but this has no biological
significance in general. It would be useful to bound e.g. the sum
of the absolute difference in sequence positions, so that we can exclude
spurious alignments between sequences which bear no meaningful relationship
to each other. We can specify this constrained sequence alignment
problem as,
\begin{equation}
s_{\text{alignsumdiff}}^{*}=\min_{l\in\mathbb{L}:\sum_{\left(i,j\right)\in l}\left|i-j\right|\leq L}\underset{\left(i,j\right)\in l}{\sum}w\left(i,j\right),\label{eq:seq-align-cons}
\end{equation}
which is in the form of (\ref{eq:constrained-semiring-spec}) when
$c\left(l\right)=T$ if $\sum_{\left(i,j\right)\in l}\left|i-j\right|\leq L$
in the min-plus semiring $\mathcal{R}$.

To solve (\ref{eq:seq-align-cons}), we can set up the simple constraint
algebra $v\left(i,j\right)=\left|i-j\right|$ and $\mathcal{M}=\left(\mathbb{N},+,0\right)$
with acceptance criteria $a\left(m\right)\le L$. As this algebra
is a group, we can insert this into (\ref{eq:lifted_product-group})
to obtain,

\begin{equation}
\begin{aligned}\left(u\otimes_{\mathcal{M}}w_{\mathcal{M}}\left(i,j\right)\right)_{m} & =\begin{cases}
i_{\oplus} & m<\left|i-j\right|\\
u_{m-\left|i-j\right|}\otimes w\left(i,j\right) & \textrm{otherwise},
\end{cases}\end{aligned}
\end{equation}
which we write as $\left(u\circledast w\left(i,j\right)\right)_{m}$
for convenience. Inserting this into (\ref{eq:DP_seqalign-poly}),
we arrive at,
\begin{equation}
\begin{aligned}f_{0,0,m}^{\mathcal{S},w} & =\begin{cases}
i_{\otimes} & m=0\\
i_{\oplus} & \textrm{otherwise}
\end{cases}\\
f_{i,0,m}^{\mathcal{S},w} & =\left(f_{i-1,0}^{\mathcal{S},w}\circledast w\left(i,0\right)\right)_{m}\\
f_{0,j,m}^{\mathcal{S},w} & =\left(f_{0,j-1}^{\mathcal{S},w}\circledast w\left(0,j\right)\right)_{m}\\
f_{i,j,m}^{\mathcal{S},w} & =\left(f_{i-1,j-1}^{\mathcal{S},w}\circledast w\left(i,j\right)\right)_{m}\oplus\left(f_{i-1,j}^{\mathcal{S},w}\circledast w\left(0,j\right)\right)_{m}\oplus\left(f_{i,j-1}^{\mathcal{S},w}\circledast w\left(i,0\right)\right)_{m},
\end{aligned}
\label{eq:DP_seqalignfixsum-poly}
\end{equation}
for all $i\in\left\{ 1,2,\ldots,N\right\} $ and $j\in\left\{ 1,2,\ldots,M\right\} $.
Thus, $s_{\text{alignsumdiff}}^{*}=\pi^{\mathcal{R},a}\left(f_{N,m}^{\mathcal{R},w}\right)=\min_{m\in\mathbb{N}:m\le L}f_{N,M,m}^{\mathcal{R},w}$.

Further rearrangements of (\ref{eq:DP_seqalignfixsum-poly}) based
on case analysis are possible and may improve the readability of the
algorithm, but as they do not generally improve implementation efficiency,
we do not explore further here. The length of alignments, lying between
$\max\left(N,M\right)$ and $N+M$, should be taken into account when
choosing the acceptance function and thereby bounding the alignment
difference sum. The result is an $O\left(N\,M\,L\right)$ time complexity
algorithm for maximum sum of absolute alignment differences $L$.

Although the alignment difference sum is convenient algebraically,
another constraint which may be useful is the maximum absolute alignment
difference. Bounding this quantity gives more precise control over
the extent to which the sequences can become misaligned before the
sequences are considered not to be matched at all. To implement this
using the algebraic theory developed above, we need the constraint
algebra $v\left(i,j\right)=\left|i-j\right|$ and algebra $\mathcal{M}=\left(\left\{ 0,1,\ldots,N^{\prime}\right\} ,\max,0\right)$,
where $N^{\prime}=\max\left(N,M\right)$ is the upper bound on the
possible sequence misalignment. Because $\mathcal{M}$ is a monoid,
we need to modify the general lifted product (\ref{eq:lift-product-map})
as follows,

\begin{equation}
\begin{aligned}\left(u\otimes_{\mathcal{M}}w_{\mathcal{M}}\left(i,j\right)\right)_{m} & =\left(\bigoplus_{\substack{m^{\prime}\in\left\{ 0,1,\ldots,N^{\prime}\right\} \\
\max\left(m^{\prime},\left|i-j\right|\right)=m
}
}u_{m^{\prime}}\right)\otimes w\left(i,j\right)\end{aligned}
.
\end{equation}

Now, we need to find an explicit expression for the set $\left\{ \max\left(m^{\prime},\left|i-j\right|\right)=m\right\} $
for $m^{\prime}\in\left\{ 0,1,\ldots,N^{\prime}\right\} $. Similar
to the situation with constrained segmentations above, there are three
cases to consider:
\begin{align}
\left\{ m^{\prime}:\max\left(m^{\prime},\left|i-j\right|\right)=m\right\}  & =\begin{cases}
\left\{ m\right\}  & m>\left|i-j\right|\\
\left\{ 0,1,\ldots,m\right\}  & m=\left|i-j\right|,\\
\emptyset & m<\left|i-j\right|
\end{cases}
\end{align}
which gives rise the following general lifted product,
\begin{equation}
\begin{aligned}\left(u\otimes_{\mathcal{M}}w_{\mathcal{M}}\left(i,j\right)\right)_{m} & =\begin{cases}
u_{m}\otimes w\left(i,j\right) & m>\left|i-j\right|\\
\left(\oplus_{m^{\prime}\in\left\{ 0,1,\ldots,m\right\} }u_{m^{\prime}}\right)\otimes w\left(i,j\right) & m=\left|i-j\right|\\
i_{\oplus} & m<\left|i-j\right|
\end{cases}\end{aligned}
\end{equation}
which we also denote by $\left(u\circledast w\left(i,j\right)\right)_{m}$
for convenience. Inserting this into (\ref{eq:DP_seqalignfixsum-poly})
gives us a novel, $O\left(N\,M\,\max\left(M,N\right)\right)$ time
DP algorithm for NW sequence alignments with an (arbitrary) constraint
on the maximum absolute difference of misalignments (Figure \ref{fig:align_fwd_timings}).

\subsection{Discrete event combinations}

As a final application exposition, in many contexts, it is important
to be able to compute quantities over sequences of discrete events.
An important application from \emph{reliability engineering }is computing
the probability of a combination of components in a complex engineered
system failing, when each failure has a given probability. A common
specification for such discrete event combinations as the semiring
problem,
\begin{equation}
s_{\text{events}}^{*}=\sum_{l\in\mathbb{L}}\underset{\left(i,n\right)\in l}{\prod}w\left(i,n\right),
\end{equation}
where $\mathbb{L}$ is the set of all possible sequences of events,
each event having probability $w\left(i,n\right)$ we will focus on
the basic \emph{failure/non-failure }case where $i=0$ for survival
and $i=1$ for failure and where $j\in\left\{ 1,2,...,N\right\} $
is the size of sequences of events being considered. Thus $s_{\text{events}}^{*}$
is the total probability over all possible sequences of $N$ fail/non-fail
events. This specification is in the form of (\ref{eq:semiring-spec})
over the \emph{probability semiring} $\mathcal{P}=\left(\left[0,1\right],+,\times,0,1\right)$
for semiring mapping $w:\left(\mathbb{N}\to\mathbb{N}\right)\to\left[0,1\right]$.
For $N=2$, the set of all possible sequences of events is,
\begin{equation}
\mathbb{L}=\left\{ \left[\left(0,1\right),\left(0,2\right)\right],\left[\left(0,1\right),\left(1,2\right)\right],\left[\left(1,1\right),\left(0,2\right)\right],\left[\left(1,1\right),\left(1,2\right)\right]\right\} .
\end{equation}

There are $2^{N}$ such sequences. A simple semiring polymorphic generator
recursion for all possible sequences of fail/non-fail events, is the
following,

\begin{equation}
\begin{aligned}f_{0}^{\mathcal{S},w} & =i_{\otimes}\\
f_{n}^{\mathcal{S},w} & =f_{n-1}^{\mathcal{S},w}\otimes\left(w\left(0,n\right)\oplus w\left(1,n\right)\right)\quad\forall n\in\left\{ 1,2,\ldots,N\right\} ,
\end{aligned}
\label{eq:subs2way-poly}
\end{equation}
where $\left(0,n\right)$ represents component non-failure and $\left(1,n\right)$
component failure at event number $n$.

In practice, engineers are interested in the more complex case of
the total probability of all possible sequences where $M$ failures
in $N$ events occurs. That means we need to constrain event sequences
$l\in\mathbb{L}$ so that exactly $M$ failures, coded as $\left(1,n\right)$,
appear in each sequence. The corresponding constrained discrete event
combinations problem is,
\begin{equation}
s_{\text{eventcombs}}^{*}=\sum_{l\in\mathbb{L}:\mathrm{fails}\left(l\right)=M}\underset{\left(i,n\right)\in l}{\prod}w\left(i,n\right),
\end{equation}
which is in the form of (\ref{eq:constrained-semiring-spec}) where
$c\left(l\right)=T$ if $\left(\mathrm{fails}\left(l\right)=\sum_{\left(i,n\right)\in l}i\right)=M$.
For $M=2$ and $N=3$, we have,
\begin{equation}
\begin{aligned}\mathbb{L}^{\prime} & =\left\{ l\in\mathbb{L}:\sum_{\left(i,n\right)\in l}i=2\right\} \\
 & =\left\{ \left[\left(1,1\right),\left(1,2\right),\left(0,3\right)\right],\left[\left(1,1\right),\left(0,2\right),\left(1,3\right)\right],\left[\left(0,1\right),\left(1,2\right),\left(1,3\right)\right]\right\} .
\end{aligned}
\end{equation}

Note that this is similar to, but subtly different from, the problem
of selecting subset size as the constraint, (\ref{eq:combs-poly}).
Using our algebraic theory, we express this constraint using the simple
algebra $\mathcal{M}=\left(\mathbb{N},+,0\right)$ which adds up the
number of failures and constraint map $v\left(i,n\right)=i$ where
$i\in\left\{ 0,1\right\} $ and acceptance criteria $a\left(m\right)=T$
if $m=M$ and false otherwise. Since $\mathcal{M}$ is a group, we
insert this into (\ref{eq:lifted_product-group}) to obtain,

\begin{equation}
\left(u\otimes_{\mathcal{M}}w_{\mathcal{M}}\left(i,n\right)\right)_{m}=\begin{cases}
i_{\oplus} & m<i\\
u_{m-i}\otimes w\left(i,n\right) & \textrm{otherwise}
\end{cases}
\end{equation}
which we can then immediately insert into (\ref{eq:subs2way-poly})
to get
\begin{equation}
\begin{aligned}f_{n,m} & =\left(f_{n-1}\otimes_{\mathcal{M}}\left(w\left(0,n\right)\oplus_{\mathcal{M}}w\left(1,n\right)\right)\right)_{m}\\
 & =\left(\left(f_{n-1}\otimes_{\mathcal{M}}w\left(0,n\right)\right)\oplus_{\mathcal{M}}\left(f_{n-1}\otimes_{\mathcal{M}}w\left(1,n\right)\right)\right)_{m}\\
 & =\left(\begin{cases}
i_{\oplus} & m<0\\
f_{n-1,m}\otimes w\left(0,n\right) & \textrm{otherwise}
\end{cases}\right)\oplus\left(\begin{cases}
i_{\oplus} & m<1\\
f_{n-1,m-1}\otimes w\left(1,n\right) & \textrm{otherwise}
\end{cases}\right)\\
 & =\left(f_{n-1,m}\otimes w\left(0,n\right)\right)\oplus\left(\begin{cases}
i_{\oplus} & m=0\\
f_{n-1,m-1}\otimes w\left(1,n\right) & \textrm{otherwise}
\end{cases}\right)
\end{aligned}
\end{equation}
suppressing semiring superscripts of $f$ for clarity. This simplifies
to the following semiring polymorphic generator recursion,

\begin{equation}
\begin{aligned}f_{0,0}^{\mathcal{S},w} & =i_{\otimes}\\
f_{0,m}^{\mathcal{S},w} & =i_{\oplus}\\
f_{n,0}^{\mathcal{S},w} & =f_{n-1,0}^{\mathcal{S},w}\otimes w\left(0,n\right)\\
f_{n,m}^{\mathcal{S},w} & =\left(f_{n-1,m}^{\mathcal{S},w}\otimes w\left(0,n\right)\right)\oplus\left(f_{n-1,m-1}^{\mathcal{S},w}\otimes w\left(1,n\right)\right),
\end{aligned}
\label{eq:event-poly-combs}
\end{equation}
for all $n\in\left\{ 1,2,\ldots,N\right\} $ and $m\in\left\{ 1,2,\ldots,M\right\} $.
In the probability semiring $\mathcal{P}=\left(\left[0,1\right],+,\times,0,1\right)$
then $w\left(1,n\right)=p_{n}$ and $w\left(0,n\right)=1-p_{n}$ correspond
to probabilities of failure and non-failure on event $n$, respectively.
An application of constrained semiring fusion (\ref{eq:DP-filter-fusion})
obtains the algorithm $s_{\text{eventcombs}}^{*}=\pi^{\mathcal{P},a}\left(f_{N}^{\mathcal{P},w}\right)=f_{N,M}^{\mathcal{P},w}$.
This $O\left(N\,M\right)$ time complexity algorithm is extremely
similar to that of \citet{radke-1994} which was derived through special,
ad-hoc reasoning. Of course, being semiring polymorphic, we can put
the generator recursion (\ref{eq:event-poly-combs}) to other uses
such as determining the most probable component failure combination
(max-product semiring) or using this as a differential component in
a machine learning system (softmax semiring).

\section{Related work\label{sec:Related-work}}

Several formal approaches to DP exist in the literature, at various
levels of abstraction. The seminal work of \citet{karp_1967} is based
on representing DP recurrences as discrete sequential decision processes,
where monotonicity justifies optimizing an associated global objective
function. This framework is not polymorphic. The work of \citet{deMoor-1991}
and others \citep{bird-1996} bases an abstraction of DP on \emph{category
theory }and \emph{relations} such as inequalities, for combinatorial
problems over arbitrary \emph{algebraic data types}. It remains to
be explored how to apply this relational approach to important DP
problems which are not focussed on \emph{optimization} problems, such
as computing complete likelihoods for hidden Markov models (the forward-backward
algorithm) \citep{little-2019}, or expectations for parameter estimation
in natural language processing problems \citep{li-2009}, for which
semirings are a natural formalism.

An interesting precursor is the model of DP described in \citet{helman_1985}.
This describes restricted forms of some of the ideas which are precisely
formulated and stated in full generality here, including the key role
of the separation of computational structure from the values which
are computed, and a special kind of homomorphic map over structural
operators, into ``choice-product'' operators. It is not polymorphic.
Implicit semiring polymorphism features in DP algorithms found in
many specialized application domains, such as natural language processing
over graphs and hypergraphs \citep{goodman-1999,li-2009,huang-2008}
and more recently in differentiable algorithms for machine learning
\citep{mensch-2018}. These studies refer to special DP algorithms
and do not address the general DP algorithm derivation problem, as
we do here.

Perhaps most closely related to our approach is the \emph{semiring
filter fusion} model of \citet{emoto-2012}, which, while not explicitly
aimed at DP, covers some algorithms which our framework addresses.
It does not address the important relationship between specification
and correct algorithm. While polymorphic, it is restricted to sequential
decision processes which can be expressed as \emph{free homomorphisms}
over associative list joins. To our knowledge, this article was first
to introduce algebraic lifting, albeit lacking proof details and in
a limited form restricted to monoids, which we expand in much greater
generality and depth here. These limitations of \citet{emoto-2012}
appear to rule out non-sequential DP algorithms e.g. sequence alignment,
edit distance and dynamic time warping, and algorithms requiring constraints
based on more complex lifting algebras such as ordered subsequences.

\section{Discussion and conclusions\label{sec:Discussion-and-conclusions}}

The paper can be summarised as follows:
\begin{itemize}
\item A method is presented for deriving efficient and correct algorithms
which solve DP problems specified as semiring objectives evaluated
over combinatorial configurations. This makes use of semiring polymorphism
and shortcut fusion from constructive algorithmics.
\item We show that more complex problems with multiple, simultaneous combinatorial
constraints can be solved within the same semiring formalism through
semiring lifting, when these constraints can be given in separable
algebraic form.
\item We show broadly applicable algebraic simplifications which can substantially
improve the computational time complexity of these derived algorithms,
where access to the form of the semiring generator is available, such
as in the case of widely available, explicit DP Bellman recursions.
\item Through the use of the tupling trick, we demonstrate a problem-agnostic
alternative to backtracing in DP algorithms which does not require
any knowledge of the specific DP algorithm.
\item Finally, derivations of multiple problems across a range of application
areas are given, which show the effectiveness of our framework for
a very wide class of problems to which DP is applicable.
\end{itemize}
While we can always express DP recurrences in a general computer language,
to do so over semirings requires special programming effort and overhead.
Modern languages which generalize classical computation to various
settings such as probabilistic or weighted logic exist and it would
be interesting to see how to implement the DP framework of this paper
in those languages. For example, \emph{semiring programming} is a
proposed overarching framework which can be considered as a strict
generalization of the polymorphic recurrences presented here \citep{belle-2020},
although this work does not address algorithm derivation. Similarly,
we can view our polymorphic DP algorithms as special kinds of \emph{sum-product
function} evaluations \citep{friesen-2016}, although, as with semiring
programming, this only describes a representation framework.

Our approach to the derivation of algorithms for constrained combinatorial
problems, requires writing these constraints in separable form using
algebras such as groups, monoids or semigroups. While this is a very
broad formalism, there will be some constraints which cannot be written
in this form. Future work may be able to provide similar algebraic
derivations when the separability requirement is relaxed. Furthermore,
the precise mapping between the approach developed in this paper and
e.g. DP algorithms specified as general discrete-continuous mathematical
optimization problems, remains an open topic.

Another issue which has not been raised is that of parallel DP implementations.
Similar approaches based on constructive algorithmics demonstrate
how to produce algorithms which are inherently parallel in the \emph{MapReduce
}framework, but these rely on associative operators and are restricted
to the setting of functional recurrences over free list semiring homomorphisms
\citep{emoto-2012}. While DP algorithms derived using our framework
are not immediately parallelisable in this way, our framework does
not rule out exploiting existing inherently parallel recurrences in
the form of free list homomorphisms, and for these recurrences, the
constraint lifting algebra developed here retains this inherently
parallel structure. The drawback is that, in some cases, it may not
be possible to simplify the lifted semiring product down to constant
time complexity as in (\ref{eq:lifted_product-group}). Future investigations
may explore general DP parallelisation frameworks \citep{galil-1994}.

A limitation of our approach as developed so far, is that it does
not exploit some of the more ``advanced'' DP speed-up tricks which
have been developed for special situations. A particular example of
this is the situation where the  mapping function $w$ in segmentation
problems satisfies a special concavity/convexity property \citep{yao-1980},
enabling a reduction in computational complexity from $O\left(N^{2}\right)$
to $O\left(N\log N\right)$. It will be interesting future work to
attempt to incorporate this and other tricks, in our framework.

Another limitation of our approach is that it does not provide a way
to derive semiring polymorphic Bellman sequential decision process
(SDP) recursions for arbitrary combinatorial generators, we must rely
upon the existence of such recursions for specific combinatorial objects.
A better approach would be to be able to derive these semiring-structured
SDPs from specification, which is addressed to a certain extent in
the constructive algorithmics literature \citep{bird-1996} and would
be a valuable direction for future research leading from this paper.

As we hope we have been able to persuade, semiring polymorphism is
not an abstract curiosity: it is an extremely useful tool for DP algorithm
derivation, as it is in many other areas of computing \citep{belle-2020,friesen-2016,goodman-1999,huang-2008,pachter-2005,mensch-2018,Sniedovich-2011}.
It offers a simple route to deriving correct DP algorithms from specifications
and quantifying computational complexity and deriving novel algorithms
in a simple, modular way through semiring lifting. It plays a central
role in clarifying what we understand to be an essential conceptual
principle of DP, which is the separation of combinatorial structure,
combinatorial constraint and value computation.

\section*{Appendix A: Proof of DP semiring fusion\label{sec:AppendixA}}

\addcontentsline{toc}{section}{Appendix A: Proof of DP semiring fusion}We
use the automated \emph{free theorem} generator Haskell package \citep{boehme-2021}
to prove (\ref{eq:DP-semiring-fusion}). Assume that the function
$f$ is implemented in some \emph{pure, lazy functional language}
(a language without side effects and without the empty type). The
type of the parameters of $f$ consists of, respectively, two binary
operators $\oplus,\otimes:\mathbb{S}\to\mathbb{S}\to\mathbb{S}$,
the mapping function $w:\mathbb{X}\to\mathbb{S}$ where $\mathbb{X}$
is an arbitrary type, and the constants $i_{\oplus},i_{\otimes}:\mathbb{S}$,
and produces a result of type $\mathbb{S}$,

\begin{equation}
f:\left(\mathbb{S}\to\mathbb{S}\to\mathbb{S}\right)\to\left(\mathbb{S}\to\mathbb{S}\to\mathbb{S}\right)\to\mathbb{S}\to\mathbb{S}\to\left(\mathbb{X}\to\mathbb{S}\right)\to\mathbb{S},\label{eq:DP_recursion_type}
\end{equation}
where $\mathbb{S}$ is an arbitrary type. According to Wadler's free
theorem \citep{wadler-1989}, this type declaration above implies
the following theorem.
\begin{thm}
For two algebraic structures $\mathcal{S}=\left(\mathbb{S},\oplus,\otimes,i_{\oplus},i_{\otimes}\right)$
and $\mathcal{S}^{\prime}=\left(\mathbb{S}^{\prime},\oplus^{\prime},\otimes^{\prime},i_{\oplus^{\prime}},i_{\otimes^{\prime}}\right)$,
assume $\mathbb{S},\mathbb{S}^{\prime}$ are arbitrary types and function
$g:\mathbb{S}^{\prime}\to\mathbb{S}$ is a map between them. Assume
also the existence of binary operators $\oplus^{\prime},\otimes^{\prime}:\mathbb{S}^{\prime}\to\mathbb{S}^{\prime}\to\mathbb{S}^{\prime}$
and $\oplus,\otimes:\mathbb{S}\to\mathbb{S}\to\mathbb{S}$, constants
$i_{\oplus^{\prime}},i_{\otimes^{\prime}}:\mathbb{S}^{\prime}$ and
$i_{\oplus},i_{\otimes}:\mathbb{S}$ and mapping functions $w^{\prime}:\mathbb{X}\to\mathbb{S}^{\prime}$,
$w:\mathbb{X}\to\mathbb{S}$. If, for all $l_{1},l_{2}:\mathbb{S}^{\prime}$
and $x:\mathbb{X}$, the function $g$ satisfies:
\begin{align*}
g\left(l_{1}\oplus^{\prime}l_{2}\right) & =g\left(l_{1}\right)\oplus g\left(l_{2}\right)\\
g\left(l_{1}\otimes^{\prime}l_{2}\right) & =g\left(l_{1}\right)\otimes g\left(l_{2}\right)\\
g\left(i_{\oplus^{\prime}}\right) & =i_{\oplus}\\
g\left(i_{\otimes^{\prime}}\right) & =i_{\otimes}\\
g\left(w^{\prime}\left(x\right)\right) & =w\left(x\right),
\end{align*}
then \emph{shortcut fusion }applies to the function $f$,

\begin{equation}
g\left(f^{\mathcal{S}',w'}\right)=f^{\mathcal{S},w}.\label{eq:DP_fusion}
\end{equation}
\end{thm}

If $\otimes$ left and right distributes over $\oplus$ and $i_{\oplus},i_{\otimes}$
are the associated identity constants, the algebraic object $\mathcal{S}=\left(\oplus,\otimes,i_{\oplus},i_{\otimes}\right)$
is a semiring. We use the shorthand $f^{\mathcal{S},w}$ to denote
$f\left(\oplus,\otimes,i_{\oplus},i_{\otimes},w\right)$ and the semiring
homomorphism $g:\mathbb{\left\{ \left[\mathbb{X}\right]\right\} }\to\mathbb{S}$
satisfying $g\left(\left\{ \left[x\right]\right\} \right)=w\left(x\right)$
by $g^{\mathcal{S},w}$. Theorem \ref{eq:DP-semiring-fusion} is a
corollary.
\begin{cor*}
DP semiring fusion. Given the generator semiring $\mathcal{G}=\left(\left\{ \left[\mathbb{X}\right]\right\} ,\cup,\circ,\emptyset,\left\{ \left[\,\right]\right\} \right)$
with the mapping function $w^{\prime}\left(x\right)=\left\{ \left[x\right]\right\} $
for all $x\in\mathbb{X}$, and another, arbitrary semiring $\mathcal{S}$
with mapping function $w:\mathbb{X}\to\mathbb{S}$, if there exists
a homomorphism $g^{\mathcal{S},w}$ mapping $\mathcal{G}\to\mathcal{S}$
which additionally satisfies $g\left(\left\{ \left[x\right]\right\} \right)=w\left(x\right)$
for all $x:\left[\mathbb{X}\right]$, then for a function $f$ with
type given in (\ref{eq:DP_recursion_type}):
\begin{equation}
g^{\mathcal{S},w}\left(f^{\mathcal{G},w^{\prime}}\right)=f^{\mathcal{S},w}.
\end{equation}
\end{cor*}

\section*{Appendix B: Constraint lifting proofs\label{sec:AppendixB}}

\addcontentsline{toc}{section}{Appendix B: Constraint lifting proofs}This
section is a generalization of the arguments given in \citet{emoto-2012},
whilst providing and clarifying essential proof details missing from
that work. The formulation of DP constraints as single operator algebras
over finite sets requires the use of (semiring) \emph{lifting }or
\emph{formal sums} as a structural tool for deriving DP constrained
fusion. This also provides a definition of the \emph{lifted semiring}
$\mathcal{S}\mathcal{M}=\left(\mathbb{M}\to\mathbb{S},\oplus_{\mathcal{M}},\otimes_{\mathcal{M}},i_{\oplus_{\mathcal{M}}},i_{\otimes_{\mathcal{M}}}\right)$.

Given a semiring $\mathcal{S}=\left(\mathbb{S},\oplus,\otimes,i_{\oplus},i_{\otimes}\right)$
and constraint algebra $\mathcal{M}=\left(\mathbb{M},\odot,i_{\odot}\right)$,
define semiring-valued formal sums $x:\mathbb{M}\to\mathbb{S}$ as
objects indicating that there are $x_{m}:\mathbb{S}$ ``occurrences''
of the element $m:\mathbb{M}$. By convention, elements $x_{m}$ taking
the value $i_{\oplus}$ are not listed. Accordingly, when two such
formal sums are added, the summation acts much like vector addition
in the semiring

\begin{equation}
\begin{aligned}\left(x+y\right)_{m} & =x_{m}\oplus y_{m}\end{aligned}
,
\end{equation}
for all $x,y:\mathbb{M}\to\mathbb{S}$. We take this to define the
lifted semiring sum $x\oplus_{\mathcal{M}}y$ elementwise, $\left(x\oplus_{\mathcal{M}}y\right)_{m}=x_{m}\oplus y_{m}$
for all $m:\mathbb{M}$. Clearly, this inherits all the properties
of $\oplus$, including commutativity and idempotency. The left/right
identity constant satisfying $x\oplus_{\mathcal{M}}i_{\oplus_{\mathcal{M}}}=i_{\oplus_{\mathcal{M}}}\oplus_{\mathcal{M}}x=x$
is just $\left(i_{\oplus_{\mathcal{M}}}\right)_{m}=i_{\oplus}$.

Next, we describe the generic \emph{change of variables }(pushforward)
formula for such formal sums. Consider an arbitrary function $f:\mathbb{M}\to\mathbb{M}$
acting to transform values from the algebra $\mathcal{M}$. We can
ask what happens to a lifted semiring object $x:\mathbb{M}\to\mathbb{S}$
under this transformation. To do this, construct the \emph{product
}semiring object on $\mathbb{M}\to\mathbb{M}\to\mathbb{S}$,
\begin{equation}
\begin{aligned}x_{m_{1},m_{2}} & =x_{m_{1}}\otimes\delta_{m_{2},f\left(m_{1}\right)}\end{aligned}
,
\end{equation}
where the lifted semiring \emph{unit function} $\delta_{m}:\mathbb{M}\to\mathbb{S}$
is defined as
\begin{equation}
\delta_{m,m^{\prime}}=\begin{cases}
i_{\otimes} & m^{\prime}=m\\
i_{\oplus} & \textrm{otherwise}.
\end{cases}
\end{equation}

Then we can ``marginalize out'' the original variable to arrive at
the change of variables formula (familiar to probability theory),
\begin{equation}
\begin{aligned}x_{m_{2}} & =\bigoplus_{m_{1}:\mathbb{M}}x_{m_{1}}\otimes\delta_{m_{2},f\left(m_{1}\right)}\\
 & =\bigoplus_{\substack{m_{1}:\mathbb{M};m_{2}=f\left(m_{1}\right)}
}x_{m_{1}}\\
 & =x_{f^{-1}\left(m_{2}\right)},
\end{aligned}
\end{equation}
where the last step holds if $f$ has a unique inverse.

A key step in proving the constrained version of DP semiring fusion,
is to be able to fuse the composition of the constraint filtering
followed by a semiring homomorphism, into a single semiring homomorphism.
To do this, we will lift the constraint filtering over the set $\mathbb{M}$.
Assume the shorthand $g^{\prime}:\left\{ \left[\mathbb{X}\right]\right\} \to\mathbb{M}\to\mathbb{S}$
and $\phi_{m}^{\prime}=\phi^{\mathcal{M},v,\delta_{m}}$ where the
acceptance function $\delta_{m}\left(m^{\prime}\right)=T$ if $m^{\prime}=m$
and $F$ otherwise. We write

\begin{equation}
g_{m}^{\prime}\left(x\right)=\left(g^{\mathcal{S},w}\cdot\phi^{\mathcal{M},v,\delta_{m}}\right)\left(x\right).
\end{equation}

Thus, $g_{m}^{\prime}\left(x\right)$ denotes the result of first
filtering the set of lists $x$ to retain any lists on which the constraint
evaluates to $m$, and then applying the homomorphism $g^{\mathcal{S},w}$
to the remaining lists. Now, for $g_{m}^{\prime}$ to be a semiring
homomorphism, it must preserve semiring structure. For it to be consistent
with the filtering, it must also preserve the action of the filtering
under $\phi^{\mathcal{M},v,\delta_{m}}$.

Turning to the semiring sum, we have,
\begin{equation}
\begin{aligned}g_{m}^{\prime}\left(x\cup y\right) & =g^{\mathcal{S},w}\cdot\phi_{m}^{\prime}\left(x\cup y\right)\\
 & =g^{\mathcal{S},w}\cdot\left(\phi_{m}^{\prime}\left(x\right)\cup\phi_{m}^{\prime}\left(y\right)\right)\\
 & =\left(g^{\mathcal{S},w}\cdot\phi_{m}^{\prime}\right)\left(x\right)\oplus\left(g^{\mathcal{S},w}\cdot\phi_{m}^{\prime}\right)\left(y\right)\\
 & =g_{m}^{\prime}\left(x\right)\oplus g_{m}^{\prime}\left(y\right).
\end{aligned}
\label{eq:lift_homo_sum}
\end{equation}

To explain the second step: note that forming the union of sets of
lists has no effect on the computation of the constraint value which
determines the result of filtering. Thus, the union of sets of lists
is invariant under the action of the filter. The third step follows
because $g^{\mathcal{S},w}$ is a semiring homomorphism.

Somewhat more complex is the semiring product, for which we have
\begin{equation}
\begin{aligned}g_{m}^{\prime}\left(x\circ y\right) & =\left(g^{\mathcal{S},w}\cdot\phi_{m}^{\prime}\right)\left(x\circ y\right)\\
 & =g^{\mathcal{S},w}\left(\bigcup_{\forall m^{\prime},m^{\prime\prime}\in\mathbb{M}:m^{\prime}\odot m^{\prime\prime}=m}\left(\phi_{m^{\prime}}^{\prime}\left(x\right)\circ\phi_{m^{\prime\prime}}^{\prime}\left(y\right)\right)\right)\\
 & =\bigoplus_{\forall m^{\prime},m^{\prime\prime}\in\mathbb{M}:m^{\prime}\odot m^{\prime\prime}=m}g^{\mathcal{S},w}\cdot\left(\phi_{m^{\prime}}^{\prime}\left(x\right)\circ\phi_{m^{\prime\prime}}^{\prime}\left(y\right)\right)\\
 & =\bigoplus_{\forall m^{\prime},m^{\prime\prime}\in\mathbb{M}:m^{\prime}\odot m^{\prime\prime}=m}\left(g^{\mathcal{S},w}\cdot\phi_{m^{\prime}}^{\prime}\right)\left(x\right)\otimes\left(g^{\mathcal{S},w}\cdot\phi_{m^{\prime\prime}}^{\prime}\right)\left(y\right)\\
 & =\bigoplus_{\forall m^{\prime},m^{\prime\prime}\in\mathbb{M}:m^{\prime}\odot m^{\prime\prime}=m}g_{m^{\prime}}^{\prime}\left(x\right)\otimes g_{m^{\prime\prime}}^{\prime}\left(y\right).
\end{aligned}
\label{eq:lift_homo_prod}
\end{equation}

Clearly, this motivates the definition of the lifted semiring product
as $\left(x\otimes_{\mathcal{M}}y\right)_{m}=\bigoplus_{\forall m^{\prime},m^{\prime\prime}\in\mathbb{M}:m^{\prime}\odot m^{\prime\prime}=m}x_{m^{\prime}}\otimes y_{m^{\prime\prime}}$.

The second step above deserves further explanation. We need to be
able to push the filter $\phi_{m}^{\prime}$ inside the cross-join,
which is critical to defining a semiring homomorphism. Recall that
the cross-join $x\circ y$ of two sets of lists involves joining together
each list in $x$ with each list of $y$. For general lists $l^{\prime},l^{\prime\prime}$
whose constraints evaluate to $m^{\prime}$ and $m^{\prime\prime}$
respectively, then due to the separability of the constraint algebra,
the constraint value of their join $l^{\prime}\cup l^{\prime\prime}$
is $m^{\prime}\odot m^{\prime\prime}$. If we group together into
one set $s^{\prime}$, all those lists whose constraints evaluate
to $m^{\prime}$ and into another set $s^{\prime\prime}$, all those
lists whose constraints evaluate to $m^{\prime\prime}$, then their
cross-join $s^{\prime}\circ s^{\prime\prime}$ will consist of sets
of lists, all of which have constraints evaluating to $m=m^{\prime}\odot m^{\prime\prime}$.
Finally, for a given $m$ and without further information on the properties
of $\odot$, we can find the values of $m^{\prime},m^{\prime\prime}$
such that $m^{\prime}\odot m^{\prime\prime}=m$ by exhaustively considering
all possible pairs. Clearly, if $\odot$ is specialized in some way,
particularly with regards to the existence of inverses, then this
exhaustive search can be reduced, and this is the basis of our algebraic
simplifications for special cases such as group lifting algebras.

A semiring homomorphism must map identities. For empty sets which
are the identity for $\cup$, we simply require,
\begin{equation}
g_{m}^{\prime}\left(\emptyset\right)=i_{\oplus}\quad\forall m\in\mathbb{M}.\label{eq:lift_homo_idsum}
\end{equation}

Similarly, sets of empty lists act as identities for the cross-join
operator. In this case, we must have $g_{m}^{\prime}\left(\left\{ \left[\,\right]\right\} \circ x\right)=g_{m}^{\prime}\left(x\circ\left\{ \left[\,\right]\right\} \right)=g_{m}^{\prime}\left(x\right)$.
If we set $g_{m}^{\prime}\left(\left\{ \left[\,\right]\right\} \right)=\delta_{i_{\odot},m}$,
then we have,
\begin{equation}
\begin{aligned}g_{m}^{\prime}\left(\left\{ \left[\,\right]\right\} \circ x\right) & =\bigoplus_{\forall m^{\prime},m^{\prime\prime}\in\mathbb{M}:m^{\prime}\odot m^{\prime\prime}=m}\delta_{i_{\odot},m^{\prime}}\otimes g_{m^{\prime\prime}}^{\prime}\left(x\right)\\
 & =\bigoplus_{\forall m^{\prime\prime}\in\mathbb{M}:i_{\odot}\odot m^{\prime\prime}=m}\delta_{i_{\odot},i_{\odot}}\otimes g_{m^{\prime\prime}}^{\prime}\left(x\right)\\
 & =\bigoplus_{\forall m^{\prime\prime}\in\mathbb{M}:i_{\odot}\odot m^{\prime\prime}=m}g_{m^{\prime\prime}}^{\prime}\left(x\right)\\
 & =g_{m}^{\prime}\left(x\right),
\end{aligned}
\label{eq:lift_homo_idprod}
\end{equation}
and similarly for $g_{m}^{\prime}\left(x\circ\left\{ \left[\,\right]\right\} \right)$.
This shows the lifted semiring identity to be $i_{\otimes_{\mathcal{M}}}=\delta_{i_{\odot}}$.

Finally, we need to consider the homomorphic mapping of sets with
single element configurations e.g. terms like $\left\{ \left[x\right]\right\} $.
Under the action of the filter $\phi^{\mathcal{M},v,\delta_{m}}$,
such terms are only retained if the constraint  mapping $v\left(s\right)=m$,
whereupon they contribute a value $w\left(s\right)$ to the semiring
value of the homomorphism $g^{\mathcal{S},w}$. Otherwise, they do
not contribute anything to the semiring sum. It follows that,
\begin{equation}
\begin{aligned}g_{m}^{\prime}\left(\left\{ \left[x\right]\right\} \right) & =\delta_{v\left(x\right),m}\otimes w\left(x\right)\\
 & =\begin{cases}
w\left(x\right) & m=v\left(x\right)\\
i_{\oplus} & \textrm{otherwise},
\end{cases}
\end{aligned}
\label{eq:lift_homo_single}
\end{equation}
which we write as the mapping $w_{\mathcal{M}}\left(x\right)_{m}$.
To summarize then, (\ref{eq:lift_homo_sum})-(\ref{eq:lift_homo_single})
show that $g_{m}^{\prime}$ is a semiring homomorphism performing
the lift mapping $\mathcal{G}\to\mathcal{S}\mathcal{M}$:
\begin{equation}
g^{\mathcal{S},w}\left(\phi^{\mathcal{M},v,\delta_{m}}\right)=g^{\mathcal{S}\mathcal{M},w_{\mathcal{M}}}.
\end{equation}

The next step is to reconstruct the result of DP constraint filtering
$\phi^{\mathcal{M},v,\delta_{m}}$, from the lifted result. This involves
computing the effect of the transformation $a:\mathbb{M}\to\mathbb{B}$
mapping the lifting algebra $\mathbb{M}$ into the value in $\mathbb{B}$
of the predicate $a$, on an arbitrary lifted semiring object $x:\mathbb{M}\to\mathbb{S}$.
The joint product function $\pi$ on $\mathbb{M}\times\mathbb{B}$
is written using the Boolean-semiring unit function:
\begin{equation}
\begin{aligned}\pi^{m,b} & \left(x\right)=x_{m}\otimes\delta_{b,a\left(m\right)}\\
\delta_{b,b^{\prime}} & =\begin{cases}
i_{\otimes} & b^{\prime}=b\\
i_{\oplus} & \textrm{otherwise}.
\end{cases}
\end{aligned}
\end{equation}

We then project onto the second parameter of $\pi$ to obtain,
\begin{equation}
\begin{aligned}\pi^{b}\left(x\right) & =\bigoplus_{\forall m^{\prime}\in\mathbb{M}}x_{m^{\prime}}\otimes\delta_{b,a\left(m^{\prime}\right)}\\
 & =\bigoplus_{\forall m^{\prime}\in\mathbb{M}:a\left(m^{\prime}\right)=b}x_{m^{\prime}}\\
 & =x_{a^{-1}\left(b\right)},
\end{aligned}
\end{equation}
where the last line holds if $a$ has a unique inverse. We use the
notation $\pi^{\mathcal{S},a}$ as a shorthand for $\pi^{T}$ over
the semiring $\mathcal{S}$ and the acceptance criteria $a$.

Putting everything above together, we can show the following:

\begin{equation}
\begin{aligned}g^{\mathcal{S},w}\left(\phi^{\mathcal{M},v,a}\left(f^{\mathcal{G},w^{\prime}}\right)\right) & =\pi^{\mathcal{S},a}\left(g^{\mathcal{S},w}\left(\phi^{\mathcal{M},v,a}\left(f^{\mathcal{G},w^{\prime}}\right)\right)\right)\\
 & =\pi^{\mathcal{S},a}\left(g^{\mathcal{SM},w_{\mathcal{M}}}\left(f^{\mathcal{G},w^{\prime}}\right)\right)\\
 & =\pi^{\mathcal{S},a}\left(f^{\mathcal{SM},w_{\mathcal{M}}}\right)
\end{aligned}
\end{equation}
which constitutes a proof of theorem (\ref{eq:DP-filter-fusion}).
\begin{thm}
DP semiring constrained fusion. Given the generator semiring $\mathcal{G}$
with the mapping function $w^{\prime}$, and another, arbitrary semiring
$\mathcal{S}$ with  mapping function $w$, the constraint algebra
$\mathcal{M}=\left(\mathbb{M},\odot,i_{\odot}\right)$ with constraint
mapping function $v$, acceptance criteria $a:\mathbb{M}\to\mathbb{B}$,
constraint filtering function $\phi:\left(\mathbb{M}\to\mathbb{M}\to\mathbb{M}\right)\to\mathbb{M}\to\left(\mathbb{X}\to\mathbb{M}\right)\to\left(\mathbb{M}\to\mathbb{B}\right)$
and projection function $\pi:\left(\mathbb{S}\to\mathbb{S}\to\mathbb{S}\right)\to\mathbb{S}\to\left(\mathbb{M}\to\mathbb{B}\right)\to\mathbb{S}$,
then for a function $f$ with type (\ref{eq:DP_recursion_type})\emph{:}

\begin{equation}
\begin{aligned}g^{\mathcal{S},w}\left(\phi^{\mathcal{M},v,a}\left(f^{\mathcal{G},w^{\prime}}\right)\right) & =\pi^{\mathcal{S},a}\left(f^{\mathcal{SM},w_{\mathcal{M}}}\right)\end{aligned}
.
\end{equation}
\end{thm}

\section*{Appendix C: A selection of semirings\label{sec:AppendixC}}

\addcontentsline{toc}{section}{Appendix B: A selection of semirings}A
table of some useful (numerical) semirings $\mathcal{S}=\left(\mathbb{S},\oplus,\otimes,i_{\oplus},i_{\otimes}\right)$
is given below, for more details on these and other semirings, the
book by \citet{golan-1999} is an excellent reference.

\begin{center}
\begin{tabular}{|c|>{\centering}p{3cm}|c|c|c|}
\hline 
\textbf{Name} & \textbf{Example application} & \textbf{Set $\mathbb{S}$} & \textbf{Operations $\oplus,\otimes$} & \textbf{Identities $i_{\oplus},i_{\otimes}$}\tabularnewline
\hline 
\hline 
Arithmetic & Solution counting & $\mathbb{N}$ & $+,\times$ & $0,1$\tabularnewline
\hline 
Generator & Exhaustive listing & $\left\{ \left[\mathbb{X}\right]\right\} $ & $\cup,\circ$ & $\emptyset,\left\{ \left[\,\right]\right\} $\tabularnewline
\hline 
Boolean & Solution existence & $\mathbb{B}$ & $\vee,\wedge$ & $T,F$\tabularnewline
\hline 
Arithmetic & Probabilistic likelihood & $\mathbb{R}$ & $+,\times$ & $0,1$\tabularnewline
\hline 
Tropical & Minimum negative log likelihood & $\mathbb{R}^{+}$ & $\min,+$ & $\infty,0$\tabularnewline
\hline 
Softmax & Differentiable minimum negative log likelihood & $\mathbb{R}^{+}$ & $-\ln\left(e^{-x}+e^{-y}\right),+$ & $\infty,0$\tabularnewline
\hline 
Viterbi & Minimum negative log likelihood with optimal solution & $\mathbb{R}^{+}\times\left\{ \mathbb{R}^{+}\right\} $ & $\left(\min,\arg\min\right),\left(+,\cup\right)$ & $\left(\infty,\emptyset\right),\left(0,\emptyset\right)$\tabularnewline
\hline 
Expectation & Expectation-maximization & $\mathbb{R}\times\mathbb{R}^{+}$ & $\left(x+y,p+q\right),\left(py+qx,pq\right)$ & $\left(0,0\right),\left(1,0\right)$\tabularnewline
\hline 
Bottleneck & Fuzzy constraint satisfaction & $\left[0,1\right]$ & $\max,\min$ & $0,1$\tabularnewline
\hline 
Relational & Database queries & $\mathbb{S}R\mathbb{S}$ & $\cup,\bowtie$ & $\emptyset,1_{R}$\tabularnewline
\hline 
\end{tabular}
\par\end{center}

\section*{Appendix D: Some useful constraint algebras\label{sec:AppendixD}}

\addcontentsline{toc}{section}{Appendix C: Some useful constraint algebras}In
this section we list some useful example constraints and simplified
expressions for the resulting lifted semiring products, see (\ref{eq:lift_product-group}),
along with simplified expressions for the product against the lifted
single value, see (\ref{eq:lifted_product-group}).

\begin{center}
\begin{tabular}{|>{\centering}m{2cm}|>{\centering}m{3.5cm}|>{\centering}p{4.5cm}|>{\centering}p{6cm}|}
\hline 
\textbf{Example application} & \textbf{Algebra $\mathcal{M}=\left(\mathbb{M},\odot,i_{\odot}\right)$} & $\left(x\otimes_{\mathcal{M}}y\right)_{m}$ (\ref{eq:lift_product-group}) & $\left(u\otimes_{\mathcal{M}}w_{\mathcal{M}}\left(x\right)\right)_{m}$
(\ref{eq:lifted_product-group})\tabularnewline
\hline 
\hline 
Subset size & $\left(\mathbb{N},+,0\right)$ & {\scriptsize{}$\bigoplus_{m^{\prime}:\mathbb{N}}\left(x_{m^{\prime}}\otimes y_{m-m^{\prime}}\right)$} & {\scriptsize{}$\begin{cases}
i_{\oplus} & m<v\left(x\right)\\
u_{m-v\left(x\right)}\otimes w\left(x\right) & \textrm{otherwise}
\end{cases}$}\tabularnewline
\hline 
Minimum count & 
\[
\left(\left\{ 1,\ldots,M\right\} ,\min,M\right)
\]
 & {\scriptsize{}$\begin{array}{c}
\bigoplus_{m^{\prime}=m}^{M}\left(x_{m^{\prime}}\otimes y_{m}\right)\oplus\\
\bigoplus_{m^{\prime}=m+1}^{M}\left(x_{m}\otimes y_{m^{\prime}}\right)
\end{array}$} & {\scriptsize{}$\begin{cases}
u_{m}\otimes w\left(x\right) & m<v\left(x\right)\\
\left(\oplus_{m^{\prime}=m}^{M}u_{m^{\prime}}\right)\otimes w\left(x\right) & m=v\left(x\right)\\
i_{\oplus} & m>v\left(x\right)
\end{cases}$}\tabularnewline
\hline 
Maximum count & $\left(\left\{ 1,\ldots,M\right\} ,\max,0\right)$ & {\scriptsize{}$\begin{array}{c}
\bigoplus_{m^{\prime}=1}^{m-1}\left(x_{m^{\prime}}\otimes y_{m}\right)\oplus\\
\bigoplus_{m^{\prime}=1}^{m}\left(x_{m}\otimes y_{m^{\prime}}\right)
\end{array}$} & {\scriptsize{}$\begin{cases}
u_{m}\otimes w\left(x\right) & m>v\left(x\right)\\
\left(\oplus_{m^{\prime}=1}^{m}u_{m^{\prime}}\right)\otimes w\left(x\right) & m=v\left(x\right)\\
i_{\oplus} & m<v\left(x\right)
\end{cases}$}\tabularnewline
\hline 
Absolute difference & 
\begin{align*}
\left(\left\{ 1,\ldots,M\right\} ,\right.\\
\left.\left|x-y\right|,0\right)
\end{align*}
 & {\scriptsize{}$\begin{array}{c}
\bigoplus_{m^{\prime}=m+1}^{M-1}\left(x_{m^{\prime}}\otimes y_{m^{\prime}-m}\right)\oplus\\
\bigoplus_{m^{\prime}=1}^{M-m-1}\left(x_{m^{\prime}}\otimes y_{m^{\prime}+m}\right)
\end{array}$} & {\scriptsize{}$\begin{array}{c}
\left(\begin{cases}
i_{\oplus} & m>v\left(x\right)-1\\
u_{m-v\left(x\right)}\otimes w\left(x\right) & \textrm{otherwise}
\end{cases}\right)\oplus\\
\left(\begin{cases}
i_{\oplus} & m>M-v\left(x\right)\\
u_{m+v\left(x\right)}\otimes w\left(x\right) & \textrm{otherwise}
\end{cases}\right)
\end{array}$}\tabularnewline
\hline 
Existence & $\left(\mathbb{B},\vee,F\right)$ & {\scriptsize{}$\begin{cases}
x_{F}\otimes y_{F} & m=F\\
\left(x_{F}\otimes y_{T}\right)\oplus\left(x_{T}\otimes y_{F}\right) & m=T\\
\oplus\left(x_{T}\otimes y_{T}\right)
\end{cases}$} & {\scriptsize{}$\begin{cases}
u_{m}\otimes w\left(x\right) & v\left(x\right)=F\\
\left(u_{F}\oplus u_{T}\right)\otimes w\left(x\right) & \left(m=T\right)\wedge\left(v\left(x\right)=T\right)\\
i_{\oplus} & \left(m=F\right)\wedge\left(v\left(x\right)=T\right)
\end{cases}$}\tabularnewline
\hline 
For all & $\left(\mathbb{B},\wedge,T\right)$ & {\scriptsize{}$\begin{cases}
\left(x_{F}\otimes y_{F}\right)\oplus\left(x_{T}\otimes y_{F}\right) & m=F\\
\oplus\left(x_{F}\otimes y_{T}\right)\\
x_{T}\otimes y_{T} & m=T
\end{cases}$} & {\scriptsize{}$\begin{cases}
u_{m}\otimes w\left(x\right) & v\left(x\right)=T\\
\left(u_{F}\oplus u_{T}\right)\otimes w\left(x\right) & \left(m=F\right)\wedge\left(v\left(x\right)=F\right)\\
i_{\oplus} & \left(m=T\right)\wedge\left(v\left(x\right)=F\right)
\end{cases}$}\tabularnewline
\hline 
Sequential-value ordering & 
\begin{align*}
\left(\left(\mathbb{N},\mathbb{R}\right),\right.\\
\preceq,\left.z_{\preceq}\right)
\end{align*}
 & {\scriptsize{}$\bigoplus_{m^{\prime}\in\mathbb{M}:m^{\prime}\preceq m}\left(x_{m^{\prime}}\otimes y_{m}\right)$} & {\scriptsize{}$\begin{cases}
\left(\oplus_{m^{\prime}\in\mathbb{M}:m^{\prime}\preceq m}u_{m^{\prime}}\right)\otimes w\left(x\right) & m=v\left(x\right)\\
i_{\oplus} & \textrm{otherwise}
\end{cases}$}\tabularnewline
\hline 
\end{tabular}
\par\end{center}

\bibliographystyle{plainnat}
\bibliography{algebraic_dp_little_2024_v22}

\end{document}